\documentclass[prl,twocolumn,showpacs,amsmath,amssymb,superscriptaddress]{revtex4-1}

\usepackage{graphicx}
\usepackage{dcolumn}
\usepackage{bm}     
\usepackage{bbm}
\usepackage{multirow}
\usepackage{booktabs}
\usepackage{amsmath}
\usepackage{braket}
\usepackage{appendix}
\usepackage{float}
\usepackage{xcolor}
\usepackage{physics}
\usepackage{siunitx}
\usepackage{amssymb}
\usepackage{array}
\usepackage{hyperref}
\newcommand{\bb}[1]{\mathbb{#1}}

\newcommand{\pgap}{\pi\text{-gap}}

\newcommand{\kk}{\bm{k}}

\begin{document}
	\title{Dynamic bulk-boundary correspondence for anomalous Floquet topology}
	
	\author{DinhDuy Vu}
	\affiliation{Condensed Matter Theory Center and Joint Quantum Institute, Department of Physics, University of Maryland, College Park, Maryland 20742, USA}
	
	\begin{abstract}
		Periodically driven systems with internal and spatial symmetries can exhibit a variety of anomalous boundary behaviors at both the zero and $\pi$ quasienergies despite the trivial bulk Floquet bands. These phenomena are called anomalous Floquet topology (AFT) as they are unconnected from their static counterpart, emerging from the winding of the time evolution unitary rather than the bulk Floquet bands at the end of the driving period. In this paper, we systematically derive the first and inversion-symmetric second-order AFT bulk-boundary correspondence for Altland-Zirnbauer (AZ) classes BDI, D, DIII, AII. For each AZ class, we start a dimensional hierarchy with a parent dimension having $\bb{Z}$ classification, then use it as an interpolating map to classify the lower-dimensional descendants. From the Atiyah-Hirzebruch spectral sequence (AHSS), we identify the subspace that contains topological information and faithfully derive the AFT bulk-boundary correspondence for both the parent and descendants. Our theory provides analytic tools for out-of-equilibrium topological phenomena.
	\end{abstract}
	
	\maketitle
	
	\section{Introduction}
	Topological phases of matter are characterized by robust gapless boundary states arising from non-trivial bulk bands. Regarding non-interacting gapped Hamiltonians with particle-hole, time-reversal and chiral symmetries (PHS, TRS, and CS), the classification for topological phases has been explored exhaustively by K-theory and Bott periodicity, resulting in the well-known ten-fold periodic table \cite{Kitaev2009,Altland1997}. The internal symmetries generally protect $(d-1)$D gapless boundary modes in a $d-$D system, which we refer to as first-order topology. The topological invariants as well as bulk-boundary correspondence for this case have been studied extensively with experimentally relevant examples including the TKNN number for 2D class A Chern insulators \cite{Thouless1982}, the Pfaffian parity for Kitaev chains \cite{Kitaev2001}, or the $\bb{Z}_2$ order for quantum spin Hall insulators in two \cite{Kane2005,Bernevig2006} and three dimensions \cite{Fu2007,Fu2007b,Moore2007,Roy2009}. Interestingly, when spatial symmetries are present, the topological classification is significantly enriched with a plethora of robust higher-order $(d-n)$D boundary modes (here $n>1$), known as higher-order topology \cite{Fu2011,Benalcazar2017,Song2017,Langbehn2017,Khalaf2018,Ortix2018,Po2017,Ono2018,Khalaf2018b, Kruthoff2017}. Much effort has been made to enumerate higher-order topological phases and discover their bulk-boundary correspondences, most notably the systematic construction of symmetry indicators \cite{Po2017,Ono2018,Khalaf2018b}.  
	
	Topological phenomena also manifest in out-of-equilibrium systems where the time dimension provides an extra tuning parameter. Experimentally, it is most common to introduce the time dependence via applying a periodic action upon an electronic \cite{Yao2007,Inoue2010,Lindner2011,Perez2014,Torres2014}, photonic \cite{Rechtsman2013,Maczewsky2017} or cold-atom\cite{Jotzu2014,Jim2015,Yang2018} system. With $T$ being the period of the driving Hamiltonian, the system dynamics is then described by a time evolution unitary
	\begin{equation}
	U(\kk,T) = \mathcal{T}e^{-i\int_0^T H(\kk,t)dt}.
	\end{equation}
	A straightforward way to analyze the time evolution unitary is to map it into a time-independent gapped Hamiltonian, called Floquet Hamiltonian, given by $H_F(\kk)=i\ln_{\epsilon} U(\kk,T)$ with $\epsilon$ being the branch cut of the complex logarithm. For this operation to be well-defined, $U(\kk,T)$ must be gapped around the branch cut, i.e. no states sit at the branch cut. Suppose the time-dependent Hamiltonian is decomposed into the equilibrium part and the oscillating driving field $H(t)=H_0 + V(t)$ such that $T^{-1}\int_0^T H(t)dt = H_0$, the effective Floquet Hamiltonian $H_F$, in principle, can be topologically distinct from the equilibrium analog $H_0$. This provides a way to realize topological phenomena by driving an otherwise trivial system. Such effects are grouped into the Floquet topology. However, there are cases that the non-trivial topology emerges even when the Floquet Hamiltonian is trivial as a result of the evolution unitary dynamics, known as anomalous Floquet topology (AFT) \cite{Kitagawa2010,Jiang2011,Rudner2013,Nathan2015,Zhu2020}.
	
	It is instructive to see how one time evolution can be topologically distinct from another. An adiabatic deformation thus must not close the branch cut and/or break the symmetries. Throughout this paper, we choose $\epsilon = \pi$ for consistent with the PHS and thus use ``branch cut" and ``$\pgap$" interchangeably. If the phase bands (time-dependent quasienergy levels) of two unitaries cross the branch cut differently, they cannot be adiabatically deformed to each other without closing the $\pgap$ at $T$. In fact, the aforementioned AFT is the results of irremovable $\pgap$ crossings and  and uncaptured by the effective static Floquet Hamiltonian \cite{Kitagawa2010,Jiang2011,Rudner2013,Nathan2015,Zhu2020}. Therefore, studying AFT requires the knowledge of the  full time evolution history, instead of just $U(\kk,T)$ at the end of the driving period. We note that in our definition, the Floquet Hamiltonian defining the static limit has its quasienergy spectrum confined in $(-\pi/T,\pi/T)$. This condition can be, indeed, relaxed by allowing an arbitrary $2n\pi/T$ to be added to the Floquet Hamiltonian, thus the static limit itself also have non-trivial winding data. In addition, the branch cut can be chosen arbitrarily. In this case, the AFT must be associated with the quotient winding data taking into account all the ambiguity \cite{Yu2021}. In our paper, however, the final goal is to predict the boundary behavior at a specific quasienergy gap, so our more narrow definition of AFT is sufficient.
	
	The origin of AFT is contained the time evolution unitary after ``subtracting" the Floquet part. The residue, referred to as the unitary loop, has the starting and ending points being trivial (identity operator) and is unique up to homotopy. By mapping the unitary loop into a Hermitian map and applying the K-theory, it is shown that the classification for AFT is similar to the ten-fold way for static gapped insulators/superconductors \cite{Roy2017,Peng2020}. Despite being able to enumerate all possible phases, this approach does not provide the algorithm to topologically diagnose a time-evolution unitary. A general dynamic topological invariant for all dimensionalities can only be defined in complex classes A and AIII. For class A systems with an even number of spatial dimensions, the invariant is the winding number of the $(d+1)$D unitary in the spacetime BZ \cite{Nathan2015,Rudner2013,Fruchart2016}; while for chiral systems with an odd spatial dimensionality,  the dynamic bulk-boundary correspondence is built from the phase band winding number of one chiral subspace \cite{Asb2014,Fruchart2016}. Interestingly, for (2+1)D class A, the phase band winding number can be associated with the quantized Weyl charge of $\pgap$ crossings, which can be viewed as 3D Weyl points. The winding numbers decorated with spatial symmetry eigenvalues, can also diagnose higher-order AFT phases in complex AZ classes \cite{Zhang2020, Yu2021, Zhu2021}.  On the other hand, for real AZ classes, the anomalous Floquet topological invariants have been successfully derived only in isolated cases. Specifically, for (1+1)D class D systems with PHS, a Pfaffian-like invariant is constructed from the winding of phase bands across the $\pgap$ at high-symmetry momenta $k=0$ and $\pi$ \cite{Jiang2011}; for (2+1)D unitaries with TRS, it arises from the winding number of half the space-time BZ \cite{Carpentier2015}; and for (2+1)D driven systems with particle-hole and rotational symmetries, Ref.~\cite{Vu2021} obtains the first and second-order AFT bulk-boundary correspondences.  Our goal is providing the AFT analysis for other real AZ classes with different spatial dimensionalities. The main theoretical frameworks are the Atiyah-Hirzebruch spectral sequence (AHSS) and the dimensional hierarchy. Here, we briefly describe the motivation for their usage and leave the technical details for following parts.
	
	Atiyah-Hirzebruch spectral sequence is a mathematical tool to compute the K-group for a system with certain symmetries, producing the topological classification. The basic idea is to divide the configuration space (can be momentum or real space) into cells inside which each point, under all symmetry operators, is either mapped back to itself or to another cell with the same dimensionality. A $d-$dimensional configuration space is thus decomposed into 0-cells, 1-cells, $\dots$, up to $d-$cells. Each cell now has an emergent AZ class depending on which symmetries leave the cell invariant, resulting in the first-order approximation of the topological classification, known as $E_1$-pages. Then, the connections between cells, i.e., how the topological phase of one cell affects its adjacent higher-dimensional (or lower-dimensional in real space AHSS) cells, are considered iteratively until the classification converges after at most $d+1$ iterations \cite{Shiozaki2018,Stehouwer2018,Okuma2019}. 
	
	The motivation for us to apply AHSS is the similarity in the first-order topological classifications of a unitary loop and a gapped Hamiltonian having the same AZ class and spatial dimensionality, as shown in the identical period tables. Furthermore, as explained latter, the unitary loop can be redefined such that TRS and CS do not flip the time dimension \cite{Roy2017,Peng2020}, while PHS and spatial symmetries act trivially on $t$ by definition. Since both the $E_1-$pages of each cell and the inter-cell compatibility relations are unaffected by the extra time dimension, the implementation of the AHSS is identical to the case of static insulators/superconductors. We can then immediately assert that the twisted K-group of a unitary loop that classifies both the first-order and higher-order AFT in the presence of internal and spatial symmetries is isomorphic to that of a gapped Hamiltonian having the same spatial dimensionality and symmetries. For our bulk-boundary correspondence derivation, the AHSS serves another purpose as it identifies which cells of the configuration space encode AFT information, similar to the classification of gapped Hamiltonian \cite{Huang2021,Chen2021}. Physically, this tells us where robust $\pgap$ closings can exist. 
	
	Regarding the 8-fold periodic table of real K-groups, for every AZ class $n=0,\dots,7$ (corresponding to AI, $\dots$, CI), the $n-$D system always has $\bb{Z}$ classification while its immediate descendants of $n-1$ and $n-2$ dimensions have $\bb{Z}_2$ classification by the virtue of Bott periodicity (see Table.~\ref{tab1}). For AZ classes BDI, D, DIII and AII, these sequences cover most of experimentally relevant systems. Interestingly, the bulk-boundary correspondences of members of the chain $\bb{Z}\to\bb{Z}_2\to\bb{Z}_2$  are indeed related and form a dimensional hierarchy. The relation is transparent through the dimensional reduction process in which one dimension of the ascendant (high-dimensional) system is used as an interpolating parameter to classify the descendant (lower-dimensional) system \cite{Qi2008,Ryu2010}. The most important consequence is probably the axion field and its quantization in the presence of either TRS with a proper spatial symmetry or an improper one with determinant -1 \cite{Wang2010, Turner2012,Fang2012,Wieder2018,Yu2020}. In this paper, we do not attempt to formulate the dynamic version of axion field, but use the dimensional hierarchy to study how the $\pgap$-closing singularities is passed from the parent to its descendants and the subsequent boundary signatures. We find that while the $\bb{Z}$ classification of the parent is characterized by the number of singularities, only the parity (oddness or evenness) is preserved in the descendants, explaining their $\bb{Z}_2$ classification. The underlying logic of the dimensional reduction is general and can be applied to higher-order topology, allowing us to derive systematically the bulk-boundary correspondence for both the first-order and second-order AFT. 
	
	The paper is organized as follows. In Sec.~\ref{overview}, we introduce the technical details of the time evolution unitary decomposition, the AHSS, as well as the conventions used in the paper. In Secs.~\ref{firstorder} and \ref{secondorder}, we formulate the topological invariants and bulk-boundary correspondences for the first and second-order AFT in each AZ class. An example of AFT in a (2+1)D class DIII driven system is presented in Sec.~\ref{example} to demonstrate our theory. We conclude the paper and discuss some outlooks in Sec.~\ref{conclusion}.
	
	\section{Overview}\label{overview}
	\subsection{Return map and the period table}
	In principle, an arbitrary time evolution unitary can be decomposed into a unitary loop and a Floquet part, up to a homotopy. A common algorithm for such decomposition is the return map so that if $U(\kk,T)$ is gapped around the $\pi-$quasienergy, the Floquet Hamiltonian and the return map can be defined as
	\begin{equation}
	\begin{split}
	& H_F(\kk) =iT^{-1} \ln \left[U(\kk,T)\right]\\
	& R(\kk,t) = U(\kk,t)\left[U(\kk,T)\right]^{-t/T}.
	\end{split}
	\end{equation} 
	Both the $\ln$ and the complex exponent take the branch cut at $e^{i\pi}=-1$, explaining the $\pgap$ condition. Importantly, by this construction, the return map is periodic $R(\kk,0)=R(\kk,T)=\mathbbm{1}$. We summarize the action of symmetry operators on the return map as follows.
	\begin{equation}\label{eq3}
	\begin{split}
	&PR(\kk,t)P^{-1} = R^*(-\kk,t)\\
	&\theta R(\kk, t)\theta^{-1} = R^*(-\kk,T-t)\\
	&C R(\kk,t)C^{-1} = R(\kk,T-t),
	\end{split}
	\end{equation}	  
	where $P, \theta$ and $C$ are the unitary transformation corresponding to the PHS, TRS and CS respectively, while complex conjugating operator in anti-unitary symmetries contain is denoted by $\mathcal{K}$. The return map can be transformed such that the time dimension is trivial under symmetry actions \cite{Roy2017,Peng2020}
	\begin{equation}
	R_s(\kk,t)=R^{-1}\left( \kk,\frac{T-t}{2} \right)R\left(\kk,\frac{T+t}{2}\right).
	\end{equation}	 
	It is now straightforward to compute the K-group and the topological classification with the result being exactly similar to the ten-fold way of gapped Hamiltonians. We refer to Refs.~\cite{Roy2017,Peng2020} for detailed arguments, here, for convenience we present in Tab.~\ref{tab1} the periodic table for first-order AFT where the red indices indicate the dimensional hierarchies that we study in this work. We note that the symmetrized return map, despite being convenient to obtain the K-group, has a degenerate plane at $t=T/2$ because $R_s^2(\kk,T/2)=\mathbbm{1}$ so we actually derive the topological invariants based on the original return map $R(\kk,t)$. 
	
	A continuous return map, in a vicinity of any point $(\kk_0,t_0)$ in the space-time BZ, can be written as
	\begin{equation}
	R(\delta\kk, \delta t) = e^{-i\phi_n(\delta\kk,\delta t)} \ket{\psi_n(\delta \kk,\delta t)}  \bra{\psi_n(\delta \kk,\delta t)},
	\end{equation}
	where $(\delta\kk,\delta t)$ is a small displacement from the reference point $(\kk_0,t_0)$ and $\phi_n,\psi_n$ are continuous functions of the displacement. Within this expression, by imposing the continuity of $\phi_n$, we in turn relax the condition $-\pi<\phi_n\le\pi$. We can then define the instantaneous Hamiltonian
	\begin{equation}
	\begin{split}
	h(\delta\kk,\delta t) &= \left[\phi_n(\delta\kk,\delta t) -\phi_n(0,0)\right] \\
	&\times \ket{\psi_n(\delta \kk,\delta t)}  \bra{\psi_n(\delta \kk,\delta t)}.
	\end{split}
	\end{equation}
	For conciseness, the coordinates of the reference point are suppressed but can be inferred from the context. In our paper, we extensively use the concept of instantaneous Hamiltonians to diagnose the local robustness of the $\pgap$ closing points, but in the end the topological invariants are still expressed through the unitary return map.
	
	\begin{table}
		\begin{center}
			\begin{tabular}{>{\centering} m{0.04\textwidth} | >{\centering}   m{0.05\textwidth}  >{\centering}   m{0.04\textwidth} >{\centering}   m{0.04\textwidth} >{\centering}   m{0.04\textwidth} >{\centering}   m{0.04\textwidth} >{\centering}   m{0.04\textwidth} >{\centering}   m{0.04\textwidth}  >{\centering\arraybackslash} m{0.04\textwidth}}
				\hline
				& $d=0$ & 1 & 2 & 3 & 4 & 5 & 6 & 7 \\ 
				\hline
				A & $\bb{Z}$ & 0 & $\bb{Z}$ & 0 & $\bb{Z}$ & 0 & $\bb{Z}$ & 0\\
				AIII & 0 & $\bb{Z}$ & 0 & $\bb{Z}$ & 0 & $\bb{Z}$ & 0 & $\bb{Z}$\\
				\hline
				AI & $\bb{Z}$ & 0 & 0 & 0 & $\bb{Z}$ & 0 & $\bb{Z}_2$ & $\bb{Z}_2$\\
				BDI & $\bb{Z}_2$ & \textcolor{red}{ $\bb{Z}$} & 0 & 0 & 0 & $\bb{Z}$ & 0 & $\bb{Z}_2$\\
				D & $\bb{Z}_2$ & \textcolor{red}{$\bb{Z}_2$} &\textcolor{red}{$\bb{Z}$ }& 0 & 0 & 0 & $\bb{Z}$ & 0\\
				DIII & 0 & \textcolor{red}{$\bb{Z}_2$} & \textcolor{red}{$\bb{Z}_2$} & \textcolor{red}{$\bb{Z}$} & 0 & 0 & 0 & $\bb{Z}$\\
				AII & $\bb{Z}$ & 0 & \textcolor{red}{$\bb{Z}_2$} & \textcolor{red}{$\bb{Z}_2$} & \textcolor{red}{$\bb{Z}$} & 0 & 0 & 0\\
				CII & 0 & $\bb{Z}$ & 0 & $\bb{Z}_2$ & $\bb{Z}_2$ & $\bb{Z}$ & 0 & 0\\
				C & 0 & 0 & $\bb{Z}$ & 0 & $\bb{Z}_2$ & $\bb{Z}_2$ & $\bb{Z}$ & 0\\
				CI & 0 & 0 & 0 & $\bb{Z}$ & 0 & $\bb{Z}_2$ & $\bb{Z}_2$ & $\bb{Z}$\\
				\hline		
			\end{tabular}
		\end{center}
		\caption{Periodic table for the first-order AFT with $d$ denoting the number of spatial dimensions.\label{tab1}}
	\end{table}	 
	
	\subsection{Atiyah-Hizerbruch spectral sequence}
	\begin{figure}
		\includegraphics[width=0.44\textwidth]{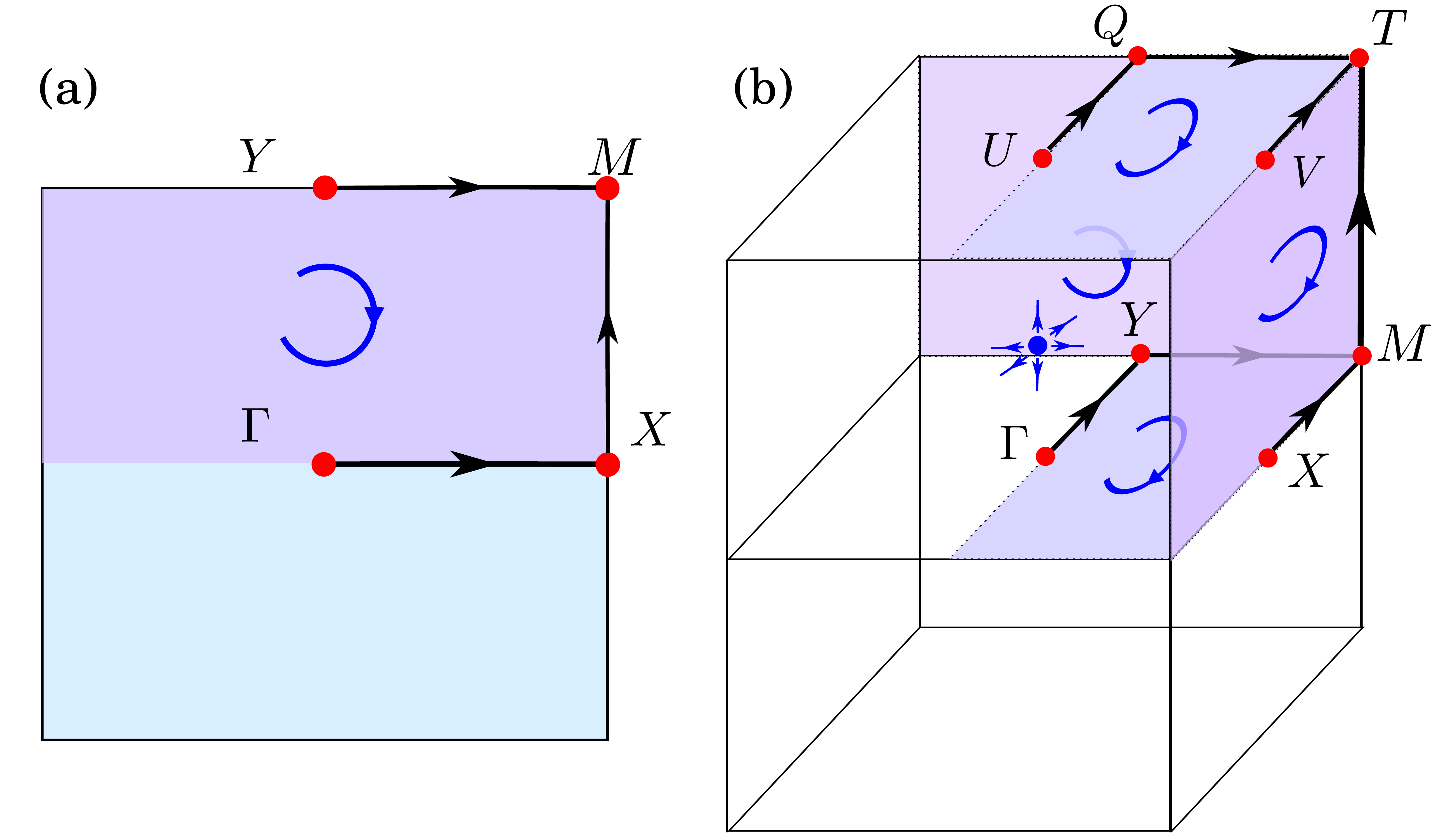}
		\caption{Cell-decomposition of 2D and 3D BZ with respect to the map $\kk\to-\kk$. The arrows denote the orientation of each cell. \label{fig2}}
	\end{figure}
	
	In this paper, we implement AHSS in the momentum-space while for the real-space classification, we refer the readers to \cite{Okuma2019}. The first step is to decompose the BZ into cells such that any symmetry operators either act trivially on the cell or map it to another cell of the same dimension. Within the set of symmetry-related cells, the orientation should be defined consistently. The K-group of each cell is determined from the emergent AZ class, i.e. how the little group of symmetry acts on the cell. If a cell is invariant under a unitary symmetry other than the identity operator, each of the irreducible representations has its own K-group, depending on whether this symmetry commutes/anti-commutes with internal symmetries. This defines the first page $E_1^{p,-(n+p)} := K^{-n}(S^p)$ of the $p-$cell having the $n$ grading defined by the emergent AZ class. Other entries $E_1^{p+r,-(n+p)}$ can be obtained by advancing $E_1^{p,-(n+p)}$ along the 8-fold (real K-group) or 2-fold (complex K-group) ways. 
	
	It is instructive to elaborate on the physical interpretation of the $E_1-$pages. Based on the definition, $E_1^{p,-(n+p)}-$pages are understood as the classification of gapped phases on $p-$dimensional space for AZ class $n$. Considering the continuous path with the time variable $t$ so that $h(\kk,t_1)$ and $h(\kk,t_2)$ being two topologically inequivalent gapped Hamiltonians, then the path $h(\kk,t)$ must have a robust gap-closing point at $t\in (t_1,t_2)$. Importantly, the classification of the gapless $h(\kk,t)$ is the same as the classification of the gapped $h(\kk)$ with $t$ as a parameter. As a result, $E_1^{p,-(n+p)}-$page is also the classification of gapless phase on the $p$-cell with the time dimension in the $(\kk,t)$ BZ. This is indeed consistent with physical origin of AFT as irremovable $\pgap$ closing points whose robust existence is shown in the AHSS. As a result, we ignore the time dimension during the cell-decomposition and proceed similarly to the case of gapped phases, but interpret the results as the classification for unitary return maps. In Fig.~\ref{fig2}, we present the cell decomposition of the 2D and 3D $\kk$-space BZ with respect to the mapping $\kk\to -\kk$. The 2D BZ is decomposed into four 0-cells, three 1-cells and one 2-cells; while the 3D BZ has eight 0-cells, seven 1-cells, four 2-cells and one 3-cells. We refer to the 0-cell as high-symmetry momentum (HsM), 1-cell as high-symmetry line (HsL) and 2-cell as high-symmetry plane (HsP).  
	
	The second step is to glue the first pages together, taking into account the action of a $p$-cell on its neighboring $(p+1)-$cell through the differential $d_1^{p,-(n+p)}:E_1^{p,-(n+p)} \to E_1^{p+1,-(n+p)}$. The first differential has physical interpretation as the extension of the gapped $p-$D cell into its adjacent $(p+1)-$D gapless cell. Taking the geometrical compatibility relation into account, the local classification is refined as $E_2^{p,-(n+p)}:=\text{Ker}\left(d_1^{p,-(n+p)}\right)/\text{Im}\left(d_1^{p-1,-(n+p)} \right)$, reminiscent of the cohomology group. Iteratively, the local K-group can be further refined by higher-order differentials linking the $p$-cell to its $(p+r)$-cell neighbors
	\begin{equation}
	d_r^{p,-(n+p)}: E_r^{p,-(n+p)} \to E_r^{p+r,-(n+p+r-1)}
	\end{equation}
	This defines the higher-order page as
	\begin{equation}
	E_{r+1}:=\text{Ker}\left(d_r^{p,-(n+p)}\right)/\text{Im}\left(d_r^{p-r,-(n+p+r-1)} \right)
	\end{equation}
	The converged local $K$-group is $E_\infty^{p,-(n+p)}:=E_{d+1}^{p,-(n+p)}$ as any differentials beyond $d$ is identically trivial. The converged pages provide the actual classification of each cell and identify the subspaces where the topological information is encoded. We note that since the differentials only depend on the spatial geometry, they also describe the connection between $p-$cells with the time dimension attached to them. As a result, the computed K-group is also applicable to AFT of return maps.
	
	\subsection{Winding number}
	An important integral that we use extensively in this work is the integer-quantized winding number of a continuous unitary gauge on an odd-dimensional hypersphere defined as
	\begin{equation}\label{winding_number}
	\nu_{2n+1}[U(\bm{v})] = \int_{\kk \in S^{2n+1}} \omega_{2n+1}[U(\bm{v};\kk)]d^{2n+1}k
	\end{equation}
	where $\bm{v}$ is the set of fixed parameters, and $\kk$ is the multi-dimensional integral variable. The winding number density is
	\begin{equation}\label{winding_density}
	\begin{split}
	\omega_{2n+1}&[U(\bm{v};\kk)]= \frac{(-1)^n n!}{(2n+1)!}\left(\frac{i}{2\pi}\right)^{n+1} \epsilon^{\alpha_1\alpha_2\dots} \\
	&\quad \times \text{Tr}\left[\left(U^{-1}\partial_{k_{\alpha_1}}U\right)\cdot \left(U^{-1}\partial_{k_{\alpha_2}} U\right)\dots \right],
	\end{split}
	\end{equation}
	where $\epsilon^{\alpha_1\alpha_2\dots}$ is the Levi-Civita symbol and $\alpha_i \in \{1,\dots,2n+1\}$. As a convention, for the winding number $\nu$, we only show the parameters of $U$, while for the winding number density $\omega$ we show both the parameters and the integral variables separated by the semicolon. 
	
	An interesting feature of the integer winding number is that despite being defined globally through the integral over the BZ, it can also be interpreted as the number of locally defined $\pgap$ closing points. 
	\begin{equation}\label{singularity}
	U(\kk) = \begin{cases}
	-\exp\left[i(k_i O_{i,j} \gamma_j)\right] & \left|\hspace{0.03in}\sum_j \left(k_iO_{i,j}\right)^2 \le \pi^2 \right. \\
	\hspace{0.5in}\mathbbm{1} &\text{  otherwise }
	\end{cases}
	\end{equation} 
	where $\gamma_i$ is the set of Clifford algebra $Cl_{2n+1,0}(\bb{R})$ generators satisfying $\{\gamma_i,\gamma_j\}=2\delta_{i,j}$, and $O$ is a real matrix (we set the degenerate quasi-energy at $\pi$ but this is not strictly required). Throughout this paper, we denote $\sigma_{1,2,3}$ as the three Pauli matrices, generators of $Cl_{3,0}$ and $\Gamma_{1,\dots,5}$ as the five 4-dimensional Gamma matrices, generators of $Cl_{5,0}$. The latter set is explicitly defined as $\Gamma_{i=1,2,3} = \sigma_3\otimes\sigma_i$, $\Gamma_4=\sigma_1\otimes\sigma_0$ and $\Gamma_5=\sigma_2\otimes \sigma_0$. We also use the short-hand notations $\Gamma_{ij} = [\Gamma_i,\Gamma_j]/(2i)$. By substituting Eq.~\eqref{singularity} into the winding number expressions~\eqref{winding_number} and \eqref{winding_density}, we show in the Appendix A that
	\begin{equation}
	\nu_{2n+1}[U] = \text{sign}\left\{\text{det}[O]\right\} = \pm 1.
	\end{equation}
	Since the winding number is quantized to integers, the structure that has $\pm 1$ winding number can be considered as the topological building block. Our classification for AFT is based on how these topological $\pgap$ crossing points are passed from the parent to the lower-dimensional descendants along the dimensional hierarchy.
	
	\section{First-order anomalous Floquet topology}\label{firstorder}
	\subsection{Class A and AIII}
	We first briefly review the dynamic bulk-boundary correspondence for complex classes A and AIII. A return map belonging to class A (AIII) has $\bb{Z}$ classification if the number of spatial dimensions is even (odd) and trivial classification otherwise. With $d=2n$, the $\bb{Z}$ index of class A return maps is simply its winding number in the odd-dimensional $(2n+1)$D space-time BZ \cite{Nathan2015,Rudner2013,Fruchart2016}. As shown in the Appendix B, the bulk $\nu_{2n+1}[R]$ gives the number of $(2n-1)$D boundary modes in the quasienergy spectrum.  
	
	For class AIII, there exist a chiral operator $C$ commuting with $R(k,T/2)$ (see Eq.~\ref{eq3})so that at $t=T/2$, the return map can be block-diagonalized according to the eigenvalue of the chiral symmetry as
	\begin{equation}\label{classAIII}
	R(\kk,T/2)=\begin{pmatrix}
	R_{C+}(\kk,T/2) & 0 \\ 0 & R_{C-}(\kk,T/2) \end{pmatrix}.
	\end{equation} 
	If $d = 2n+1$, we can defined the topological invariant as the winding number $\nu_{2n+1}[R_{C\pm}(T/2)]$ of one block \cite{Fruchart2016,Asb2014}. We show in Appendix B that this number indeed counts the boundary Dirac modes across the $\pgap$ protected by the chiral symmetry (the boundary is even-dimensional so Dirac modes can only exist in the presence of the chiral symmetry). We emphasize that $\nu_{2n+1}[R_{C+}(T/2)] = -\nu_{2n+1}[R_{C-}(T/2)]$ because $\nu_{2n+1}[R(t)]$ at fixed $t$ is quantized and continuous with $t$ but vanishes at $t=0$, making the winding number at any time slices identically zero.
	
	For real AZ classes without spatial symmetries, except for the 0-cells, the emergent AZ class is either A or AIII as both PHS and TRS flip the momentum. In fact, the $\bb{Z}$ index classifying the parent system of each dimensional hierarchy can be defined similarly to those of classes A and AIII, depending on whether the spatial dimensionality is even or odd.
	
	\subsection{Class BDI}
	From the ten-fold way periodic table~\ref{tab1}, the $\bb{Z}$ classification is realized for $d=1$. Its first descendant $d=0$ does not have a lower-dimensional boundary, so for this class we only present the classification and dynamic bulk-boundary correspondence for the parent $d=1$.  
	
	At a generic momentum $k\ne 0,\pi$, we can define an effective chiral symmetry $C=P\theta$ such that $C^2=1$ and $CR(k,t)C^{-1}=R(k,T-t)$. Similar to class AIII, the dynamic topological invariant is the 1D winding number of $C+$ block of the return map at $t=T/2$
	\begin{equation}
	\kappa_{1D}^{BDI} = \nu_1[R_{C+}(T/2)].
	\end{equation}
	To ensure that this is the correct invariant for class BDI, we check the compatibility between the defined winding number and the PHS. Since $[P\mathcal{K},C]=0$, each block $C\pm$ is PH invariant, constraining $\omega_1[R(T/2;k)]=\omega_1[R(T/2;-k)]$. Therefore, the integral over $k$ can still yield non-zero value. We will see later that the situation is completely different for class DIII where $\{P\mathcal{K},C\}=0$. Similar to class AIII, the $\bb{Z}$ index enumerate the number of CS-protected end modes pinned at the $\pgap$.

	\subsection{Class D}
	We first present the converged $E-$pages for the (2+1)D parent and the (1+1)D descendant.
	\begin{center}
		\begin{tabular}{ >{\centering}   m{0.15\textwidth}|  >{\centering}   m{0.07\textwidth}   >{\centering}   m{0.07\textwidth}   >{\centering\arraybackslash}   m{0.07\textwidth} }
			\hline
			$E_\infty^{p,-(2+p)}$ & $p=0$ & 1 & 2  \\ 
			\hline
			$d=2$ & $\bb{Z}_2^4$ & 0  & $\bb{Z}$\\  
			$d=1$ & $\bb{Z}_2^2$ & 0  &  -\\
			\hline
		\end{tabular}
	\end{center}
	Through the iterations of constructing the converged $E-$pages, the  $\bb{Z}$ index of the (2+1)D class D return map is similar to that of class A. Its bulk-boundary correspondence is also obvious, i.e. the number of anomalous gapless boundary modes. The other $\bb{Z}_2$ indices at the 0-cells can be associated with weak (protected by translational symmetries) or trivial (connected to the atomic limit) topology with the exact bulk-boundary correspondence obtained from matching the AHSS in momentum space with that in real space \cite{Huang2021}. Through the dimensional reduction, these weak indices can generate weak topology in the lower-dimensional descendants. However, in this paper, we focus on the strong topology unaffected by translational symmetries so we only study the dimensional hierarchy stemming from the strong $\bb{Z}$ index of the parent system. Another valuable information from the AHSS is that the $\bb{Z}_2$ invariant of the (1+1)D class D return map is contained in the 0-cells or HsMs.

	\subsubsection{Two-dimensional parent}
	The (2+1)D parent is characterized by the bulk 3D winding number 
	\begin{equation}
	\kappa_{2D}^{D} = \nu_3[R].
	\end{equation}
	Again, we should check the compatibility between the PHS and the invariant. The PHS imposes that $\omega_3[R(\kk,t)]=\omega_3[R(-\kk,t)]$ so the winding number integrated over the (2+1)D BZ is not trivialized. We note that a different symmetry, for example the TRS that imposes $\omega_3[R(\kk,t)]=-\omega_3[R(-\kk,-t)]$ , can forced the winding number to be zero. Similar to class A, $\nu_3[R]$ gives the number of 1D boundary modes traversing the quasienergy $\pgap$.
	
	\subsubsection{One-dimensional descendant}
	Moving to the (1+1)D class descendant $R(k,t)$, we can construct an interpolating path $\tilde{R}(k,\alpha, t)$ with $\alpha\in[0,2\pi]$ as a parameter such that
	\begin{equation}\label{1DclassD}
	\begin{split}
	\tilde{R}(k,0,t) = \mathbbm{1}, \quad \tilde{R}(k,\pi,t) = R(k,t)\\
	P\tilde{R}(k,\alpha,t)P^{-1} = \tilde{R}^*(-k,2\pi-\alpha,t)
	\end{split}
	\end{equation}
	The interpolating path realizes a (2+1)D class D return map classified by the $\bb{Z}$ winding number. There is, however, a freedom in choosing the path as the conditions \eqref{1DclassD} do not restrict to a unique function. With another $\tilde{R}'$, the difference in the winding number is given by
	\begin{equation}
	\begin{split}
	\nu_3(\tilde{R})-\nu_3(\tilde{R}') = \nu_3(g_1) + \nu_3(g_2)
	\end{split}
	\end{equation}
	with the paths $g_1$ and $g_2$ defined similar to Ref.~\cite{Qi2008} as
	\begin{equation}
	\begin{split}
	& g_1(k,\alpha,t) = \begin{cases}
	\tilde{R}(k,\alpha,t) &\text{ for } \alpha\in[0,\pi)\\
	\tilde{R}'(k,2\pi-\alpha,t) &\text{ for } \alpha\in [\pi,2\pi)\end{cases}\\
	& g_2(k,\alpha,t) = \begin{cases}
	\tilde{R}'(k,2\pi-\alpha,t) &\text{ for } \alpha\in[0,\pi)\\
	\tilde{R}(k,\alpha,t) &\text{ for } \alpha\in [\pi,2\pi)\end{cases}\\
	\end{split},
	\end{equation}
	By the PHS, it is obvious that $\omega_3[g_1(k,\alpha,t)]=\omega_3[g_2(-k,2\pi-\alpha,t)]$, leading to $\nu_3[g_1]=\nu_3[g_2]$ and thus $\nu_3[\tilde{R}]-\nu_3[\tilde{R}'] = 2\bb{Z}$, i.e. the parity of $\nu_3[\tilde{R}]$ is conserved for all symmetry-preserving choices of $\tilde{R}$. This is in fact consistent with the $\bb{Z}_2$ classification obtained for the (1+1)D class D return map from K-theory and the AHSS.  
	
\begin{figure}
		\includegraphics[width=0.42\textwidth]{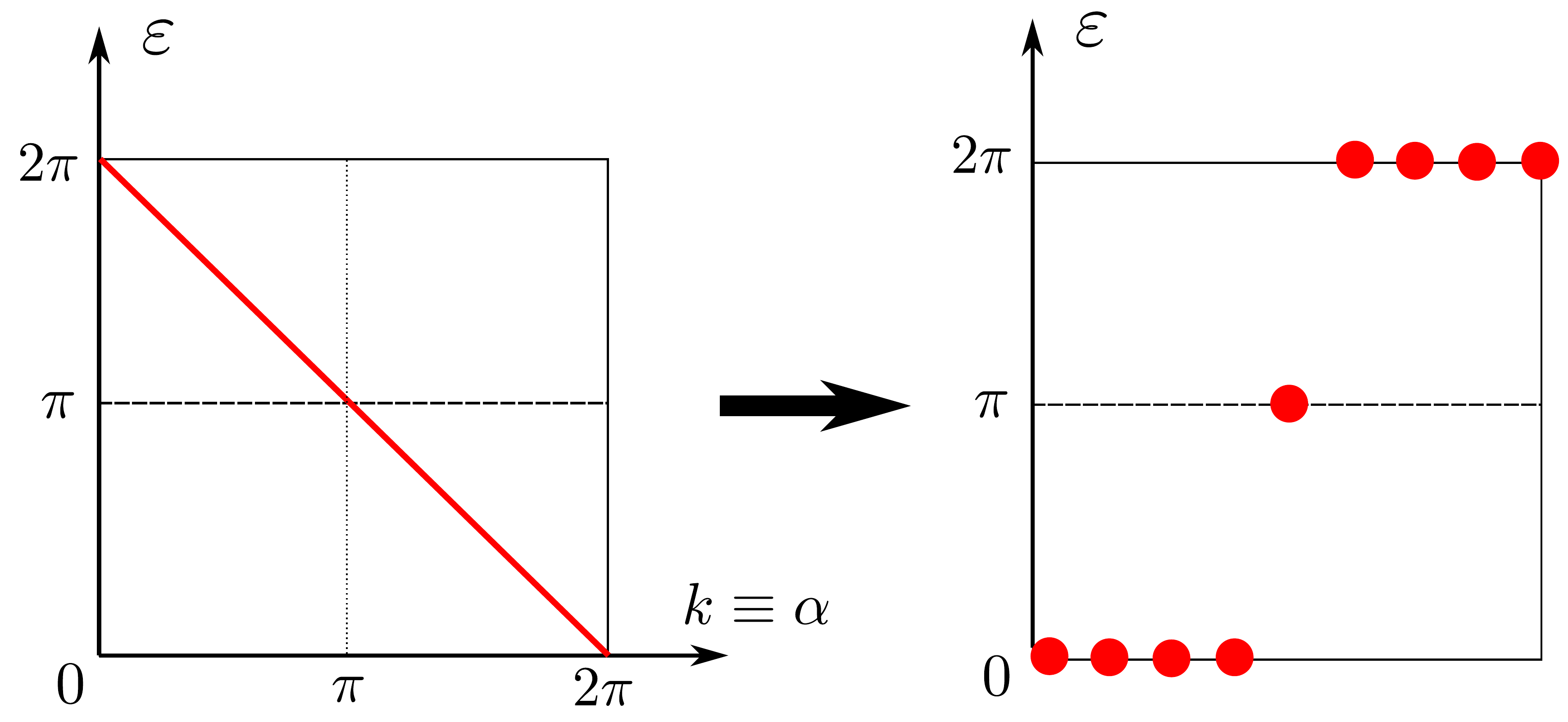}
		\caption{The passing of the anomalous boundary behavior along class D dimensional hierarchy. The first figure refers to the 1D gapless boundary mode of the (2+1)D parent at one edge. By identifying one momentum as the interpolating parameter $\alpha$ and taking a slide at $\alpha=\pi$, the non-trivial boundary behavior of the immediate descendant is obtained. The second figure shows  a single 0D majorana mode at one end of the topological (1+1)D return map pinned at $\pi-$quasienergy. \label{classDboundary}}
	\end{figure}

	As we have shown that the dimensional reduction produces the correct classification, we proceed to formulate the anomalous Floquet topological invariant based on the induction of phase band singularities in the (2+1)D interpolating path to the (1+1)D return map in interest. The winding number of a (2+1)D interpolating map is nothing but the total charge of topological phase band Weyl points. The PHS
	maps a Weyl point at a non-TR-invariant $(k,\alpha)$ to another one at $(-k,-\alpha)$ with the same charge, this pair thus does not contribute the parity of the winding number. Therefore, we only need to count the Weyl points along the $t-$axis at $k\in$ HsMs and $\alpha=\pi$. This is consistent with the AHSS argument that the $\bb{Z}_2$ classifying index for (1+1)D class D return maps is derived from information at high-symmetry momenta. Along the line, we note that $\pi_1[O(N)]=\bb{Z}_2$ where the $S^1$ can be identified the time loop at a high-symmetry momentum and the particle-hole symmetry defines a basis in which the return map is an orthogonal matrix. At a HsM, we can decompose continuously $R(\text{HsM};t)=M(t)^\dagger D(t) M(t)$ such that $D(t)=\text{Diag}[\varepsilon_1(t),\dots,\varepsilon_{N}(t),\varepsilon_1^*(t),\dots,\varepsilon_N^*(t)]$ is a smooth function of $t$, the number of $\pgap$ crossing points can be counted through the winding of one PH partner
	\begin{equation}
	\nu_1'[R(\text{HsM})] = \frac{i}{2\pi} \sum_j\int_0^T dt \varepsilon_j^*(t)\partial_t\varepsilon_j(t), 
	\end{equation} 
	where essentially only the first half of the quasienergy spectrum is used.
	There is a freedom to exchange $\varepsilon(t)$ with its PH partner $\varepsilon^*(t)$. However, this only changes $\nu_1'[R]$ by an even number, thus preserving the parity. Therefore, the total charge parity of all $\pgap$ singularities in the (2+1)D interpolating path can be expressed in the (1+1)D return map through
	\begin{equation}
	\kappa_{1D}^D = \nu_{1}'[R(0)]+\nu_1'[R(\pi)] \mod 2,
	\end{equation}
	with $\kappa_{1D}^D=0(1)$ indicates trivial (topological) AFT. The topological (1+1)D class D return maps corresponds to the interpolating map $\tilde{R}$ having an odd winding number and thus must be equipped with at least one (more precisely an odd number) gapless mode along its boundary across the quasienergy branch cut. The PHS pins the $\pgap$ crossing at either $\alpha=0$ or $\pi$, but by construction the $\alpha=0$ end is the trivial phase with opened $\pgap$ so the $\pgap$ crossing must happen at $\alpha=\pi$ as shown in Fig~\ref{classDboundary}. Therefore, the (1+1)D class D  AFT corresponds to 0D majorana modes pinned at the $\pgap$.  
	
	\subsection{Class DIII}     
	
	For class DII, we first provide the AHSS analysis for the family starting from the parent $d=3$ with $\bb{Z}$ classification
	\begin{center}
		\begin{tabular}{ >{\centering}   m{0.12\textwidth}|  >{\centering}   m{0.05\textwidth}   >{\centering}   m{0.05\textwidth} >{\centering}   m{0.05\textwidth}  >{\centering\arraybackslash}   m{0.05\textwidth} }
			\hline
			$E_\infty^{p,-(3+p)}$ & $p=0$ & 1 & 2 & 3 \\ 
			\hline
			$d=3$ & 0 & $\bb{Z}_2^7$  & 0 & $\bb{Z}$\\  
			$d=2$ & 0 & $\bb{Z}_2^3$ & 0  &  -\\
			$d=1$ & 0  & $\bb{Z}_2$   &  - & -\\
			\hline
		\end{tabular}
	\end{center}     
	The $\bb{Z}$ index of the 3-cell is identical to the that of a return map with chiral symmetry, in fact due to the existence of both particle-hole and time-reversal symmetries, the chiral symmetry can be straightforwardly defined as the product of the former two. For $d=2$ and $d=1$, the topological information is encoded in HsLs.
	
	\subsubsection{Three-dimensional parent}
	The (3+1)D class DIII unitary loop has an effective chiral symmetry $C=iP\theta$ and  can classified by the winding number of the positive-chirality block at $t=T/2$. In particular, by the eigenvalue of the chiral operator, $R(\kk,T/2)=R_{C+}(\kk,T/2)\oplus R_{C-}(\kk,T/2)$, defining the $\bb{Z}$ index
	\begin{equation}\label{DIII_3D}
	\kappa_{3D}^{DIII}=\nu_3[R_{C+}(T/2)].
	\end{equation}
	We remind that $\nu_3[R_{C+}(T/2)]=-\nu_3[R_{C-}(T/2)]$ for the same reason discussed in class AIII.
	
	At this point, it is instructive to contrast classes DIII/CI with classes BDI/CII in the (3+1)D space-time BZ. In the former case, the effective CS is $C=iP\theta$ since $(P\theta)^2=-1$, making $\{P\mathcal{K},C\}=0$, i.e. the PHS maps the $C+$ subspace to the $C-$ subspace. In the latter case, $C=P\theta$ since $(P\theta)^2=1$, so $[P\mathcal{K},C]=0$. Within the subspace $R_{C+}(T/2)$, the PHS enforces $\omega_3[R_{C+}(T/2;\kk)] = -\omega_3[R_{C+}(T/2;-\kk)]$, trivializing $\nu_3[R_{C+}(T/2)]$. Therefore, even though our earlier defined invariant only explicitly requires the CS, it is only well defined in classes DIII/CI. This CS protected winding number characterizes the number of boundary chiral 2D Dirac modes across the $\pgap$.
	
	Another way to understand the $\bb{Z}$ index is to artificially introduce an extra dimension $\beta\in [-\pi,\pi)$ such that
	\begin{equation}
	\begin{split}
	& \tilde{R}(\kk,t,0) = R(\kk,t) \\
	& C \tilde{R}(\kk, t, \beta)C^{-1} = \tilde{R}(\kk,T-t,-\beta)  \\
	& P \tilde{R}(\kk, t, \beta)P^{-1} = \tilde{R}^*(-\kk,t,\beta)  	 
	\end{split}   
	\end{equation}
	Notably, phase band Dirac modes at $(t,\beta)=(T/2,0)$ are pinned by the CS, while those off that point must exist in pairs and can be symmetrically moved away and annihilated by a choice of $\beta$ [see Fig.~\ref{DIII_diagram}(a)]. In this 5D space-time BZ, we assume the explicit form of a Dirac mode with unity charge as
	\begin{equation}\label{5D_Dirac}
	h(\delta\kk,\delta t,\beta) = \sum_{i=1}^3 \delta k_i\Gamma_i + \delta t\Gamma_4 + \beta\Gamma_5,
	\end{equation} 
	Within this basis, the only representations of the CS and PHS are $C=\Gamma_{45}$ and $P=\Gamma_{24}$. Projecting onto the subspace with $\delta t = \beta = 0$, Eq.~\eqref{5D_Dirac} describes two decoupled $\pgap$ 3D Dirac modes with opposite charges and chiral eigenvalues. This is exactly the topological invariant defined in Eq.~\ref{DIII_3D}.
	
	\subsubsection{Two-dimensional descendant}
	\begin{figure}
		\includegraphics[width=0.48\textwidth]{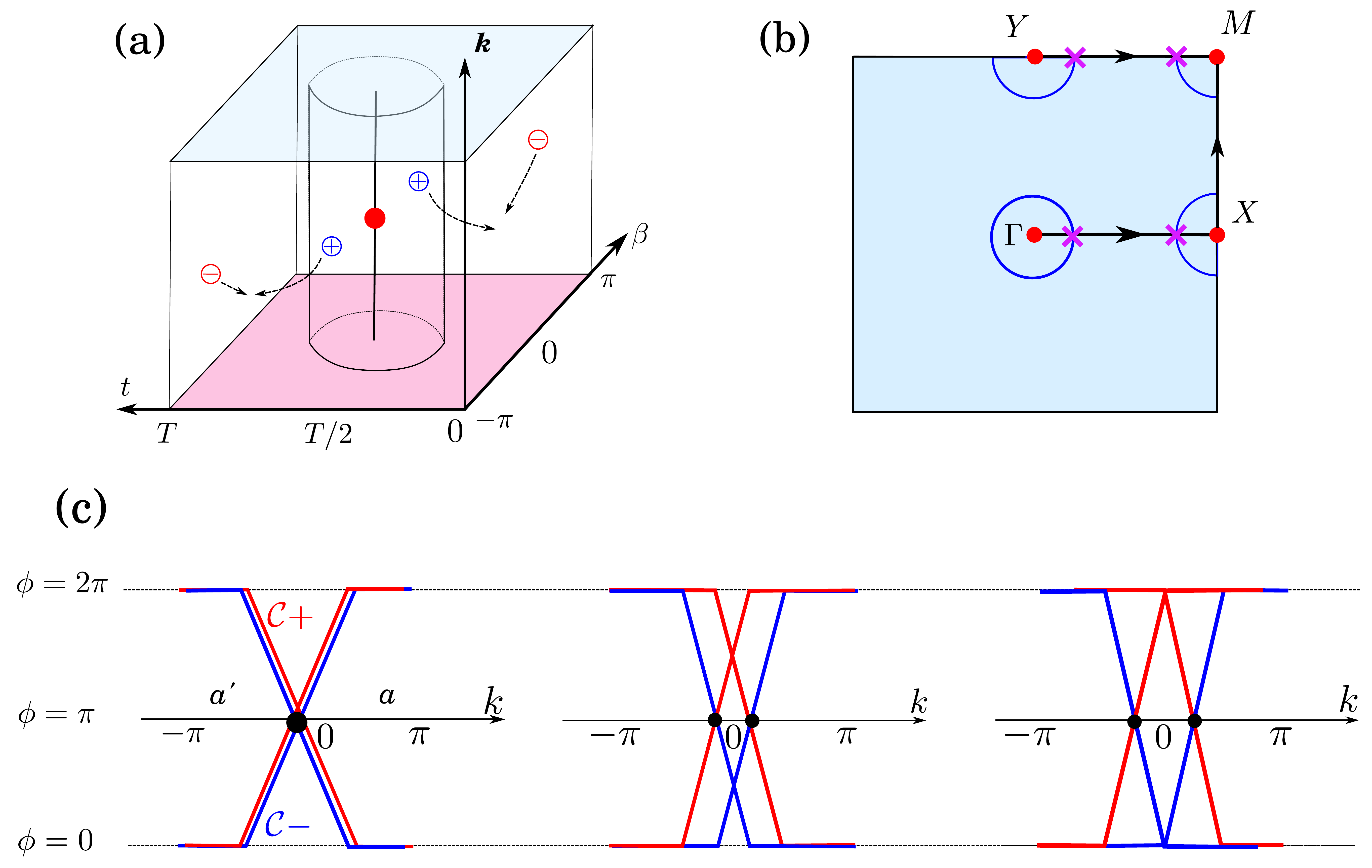}
		\caption{(a) 5D Dirac singularity in the enlarged space, the multidimensional $\kk$ is compressed into a 1D axis for visual clarity. The singularity away from $(\beta=0,t=T/2)$ has a chiral partner at $(-\beta,T-t)$, the pair can be symmetrically moved away and annihilated by the choice of $\beta$. (b) The splitting of the four-fold degenerate point at a HsM into a two-fold degenerate ring, inducing non-trivial winding number into the adjacent HsLs. (c) The adiabatic trivialization of the return map at HsM, making the winding number over half the 1D BZ well-defined.\label{DIII_diagram}}
	\end{figure}
	
	The first step in advancing along the dimensional hierarchy is to make one momentum of the parent system into a parameter in the symmetry-preserving interpolating map for the descendant. However, the action of the PHS within a chiral subspace
	\begin{equation}
	P\tilde{R}_{C+}\left(\kk,\alpha,T/2\right)P^{-1} = \tilde{R}_{C-}^*\left(-\kk,2\pi-\alpha,T/2\right),
	\end{equation} 
	is not helpful in deriving the classification because it maps one chiral subspace to another. However, from the perspective of the enlarged space with artificial dimension $\beta$,  it is more natural to study the dimensional hierarchy as we can defined the interpolating path
	$\tilde{R}(\kk,\alpha,t,\beta)$ such that
	\begin{equation}
	\begin{split}
	&\tilde{R}(\kk,0,t,0) = \mathbbm{1}, \quad \tilde{R}(\kk,\pi,t,0) = R(\kk,t)\\
	& C \tilde{R}(\kk, \alpha, t, \beta)C^{-1} = \tilde{R}(\kk,\alpha, T-t,-\beta)  \\
	& P \tilde{R}(\kk, \alpha, t, \beta)P^{-1} = \tilde{R}^*(-\kk,-\alpha, t,\beta) 
	\end{split}
	\end{equation}  
	Since the enlarged interpolating path is characterized by the singularities along the axis $(t,\beta)=(T/2,0)$, we further assume that there exist no off-axis singularities without loss of generality as explained above. The PHS imposes that $\omega_5[\tilde{R}(\kk,\alpha,t,\beta)]=\omega_5[\tilde{R}(-\kk,-\alpha,t,\beta)]$, allowing the dimensional reduction process to be implemented exactly similar to the presented class D case. As a result, $\nu_5[\tilde{R}]$ has conserved parity for all symmetry-preserving choices of the interpolating map, proving the $\bb{Z}_2$ classification of the (2+1)D class DIII return maps.
	
	The parity of the total charge $\nu_5[\tilde{R}]$ only depends on the four-fold Dirac modes at $\kk=$HsM, $\alpha=\pi$, $t=T/2$. Such a Dirac mode is described by
	\begin{equation}\label{DIII_singularity}
	h(\delta \kk, \delta t) = \delta k_1\Gamma_1+\delta k_2\Gamma_2 + \delta t \Gamma_4.
	\end{equation}
	A symmetry-preserving homogeneous term $m$ must satisfy: $\{m,h\}=0$, $[m,C] = 0$ and $\{m,P\mathcal{K}\} = 0$. The only gap-opening mass terms $\propto \Gamma_3,\Gamma_5$ are not allowed by the PHS and CS ($C=\Gamma_{45}$ and $P=\Gamma_{24}$), justifying the non-trivial classification; moreover, the terms $\propto \Gamma_4$ are not allowed either, showing that the Dirac mode is indeed pinned at $t=T/2$. Even thought the gap closing  and its associated topological charge are robust, the four-fold degenerate point is not. In fact, one can introduce a symmetry-preserving term $\lambda\Gamma_{45}$ that does not open the gap but transform the degenerate point into a degenerate ring at $\delta k_1^2+\delta k_2^2 = \lambda^2$. As shown in Fig.~\ref{DIII_diagram}(b), these nodal rings around a HsM necessarily thread though the adjacent HsL, suggesting us to look at the 1-cells for the topological invariant. This agrees with the AHSS analysis showing that the topological information is encoded in three 1-cells.
	
	We first emphasize that a 1-cell is not periodic but only half of a one-dimensional BZ so we cannot immediately evaluate any winding number. For demonstration, we restrict to a 1D k-space BZ (comprised of two related 1-cells) and consider a 4-fold degenerate point in the form of Eq.~\ref{DIII_singularity} at $k=0$, which can split into 2-fold degenerate points (the projection of the actual nodal ring) in the two adjacent 1-cells. With the $\pgap$ being opened, the return map at the two HsMs $k=0,\pi$ can be adiabatically trivialized to $\mathbbm{1}$, making the 1-cells periodic [see Fig.~\ref{DIII_diagram}(c)]. The original 4-fold degenerate point now translates to the 1D winding number of the $C+$ subspace over the closed 1-cell. Because the 1D winding number is invariant against adiabatic deformation, we can choose an explicit deformation that closes the 1-cell as
	\begin{equation}
	\begin{split}
	\bar{R}_{C+}(k) &= R_{C+}(k)\left[R_{C+}(\pi)\right]^{-\frac{k}{\pi}} \left[R_{C+}\left(0\right)\right]^{-1+\frac{k}{\pi}}.
	\end{split}
	\end{equation} 
	Here, for conciseness, we suppress the coordinates of time and perpendicular momenta, and only keep the momentum along the 1-cell in the argument.  This algorithm is similar to the one used to extract the periodic return map from the time evolution unitary. The topological invariant $\nu_1[\bar{R}_{C+}(a,T/2)]$ is now well defined. 
	
	Before moving on, we need to address two questions regarding what happens to the defined invariant if we (i) choose the other half of the one-dimensional BZ as the 1-cell and (ii) consider the $C-$ subspace instead of the $C+$. For the similar reason of the winding number continuity as presented in class AIII,
	\begin{equation*}
	\begin{split}
	& \nu_1[R_{C+}(a,T/2)] +  \nu_1[R_{C+}(a',T/2)] \\
	&\quad  + \nu_1[R_{C-}(a,T/2)]+ \nu_1[R_{C-}(a',T/2)] = 0;
	\end{split}
	\end{equation*}
	but the PHS also imposes that $\nu_1[R_{C+}(a,T/2)]=\nu_1[R_{C-}(a',T/2)]$. As a result, 
	\begin{equation}
	\begin{split}
	\nu_1[R_{C+}(a,T/2)] & = -\nu_1[R_{C+}(a',T/2)] \\
	& =-\nu_1[R_{C-}(a,T/2)].
	\end{split}
	\end{equation}
	The freedom of choices in the 1-cell and chiral subspace thus only modifies the invariant by a sign, irrelevant to its parity. Therefore, our defined topological invariant on 1-cell has a unique parity.
	
	From Fig.~\ref{DIII_diagram}(b), the classification depending on the parity of the total number of phase band 4-fold Dirac modes can be expressed as
	\begin{equation}\label{2DclassDIII}
	\begin{split}
	\kappa_{2D}^{DIII} = &\nu_1[\bar{R}_{C+}(\overline{\Gamma X},T/2)] \\
	& + \nu_1[\bar{R}_{C+}(\overline{YM},T/2)] \mod 2.
	\end{split}
	\end{equation}
	Here, $\kappa_{2D}^{DIII} = 0$ (1) corresponds to the trivial (topological) (2+1)D class AIII return maps. The boundary behavior is a slide at $\alpha=\pi$ cutting through the 2D chiral Dirac modes at the boundary of the (3+1)D parent system (see Fig.~\ref{DIII_boundary}), manifesting as a pair of 1D helical modes across the quasienergy BZ. The other two $\bb{Z}_2$ indices correspond to the weak topological phases resulted from the stacking of topological 1D chains. These weak phases are protected by the translational symmetry along $x-$ or $y-$direction.
	
	\subsubsection{One-dimensional descendant}
\begin{figure}
		\includegraphics[width=0.49\textwidth]{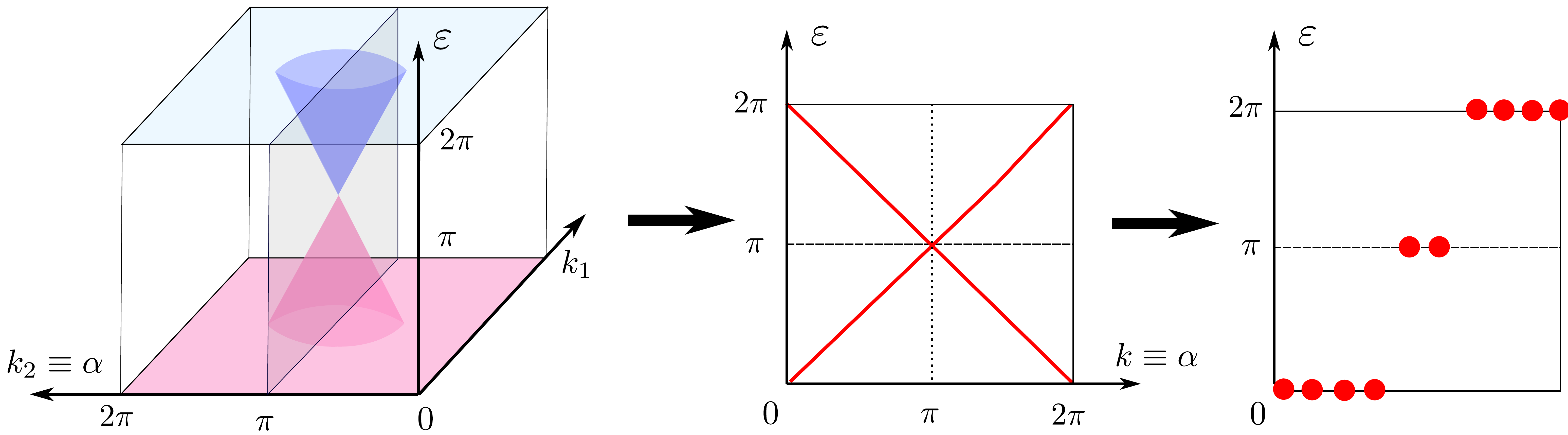}
		\caption{Anomalous boundary behavior of members in class DIII dimension hierarchy obtained by iteratively taking a slide on boundary of the ascendant. The sequence from left to right is: chiral 2D Dirac modes for the (3+1)D parent, 1D helical modes for the (2+1)D descendant, and 0D majorana Kramers pairs for the (1+1)D descendant.\label{DIII_boundary}}
	\end{figure}

	We can further advance the dimensional hierarchy and obtain the classification for the (1+1)D class DIII return map. In a similar manner, we first construct a symmetry-preserving interpolating map $\tilde{R}(k,\alpha,t)$ such that
	\begin{equation}
	\begin{split}
	&\tilde{R}(k,0,t) = \mathbbm{1},\quad \tilde{R}(k,\pi,t) = R(k,t)\\
	&P\tilde{R}(k,\alpha,t)P^{-1} = \tilde{R}^*(-k,2\pi-\alpha,t)\\
	&\theta\tilde{R}(k,\alpha,t)\theta^{-1} = \tilde{R}^*(-k,2\pi-\alpha,T-t).
	\end{split}
	\end{equation}
	The interpolating map is nothing but a (2+1)D class D return map, here we choose to identify $k_y$ with $\alpha$ so that the high-symmetry lines $(\overline{\Gamma X}, \alpha=0)$ and $(\overline{\Gamma X}, \alpha=\pi)$ correspond to $\overline{\Gamma X}$ and $\overline{YM}$ in Eq.~\eqref{2DclassDIII}. By construction, $\tilde{R}(\overline{\Gamma X},0,T/2)$ is trivial, so Eq.~\eqref{2DclassDIII} effectively provides the topological invariant for the (1+1)D class DIII return map as
	\begin{equation}
	\kappa_{1D}^{DIII} = \nu_1[\bar{R}_{C+}(\overline{\Gamma X},T/2)].
	\end{equation}
	Here, $\kappa_{1D}^{DIII} = 0$ (1) indicates the trivial (topological) phase with the boundary of the topological phase hosting majorana Kramers pairs in the $\pgap$ which can be thought of as a slide of the helical boundary modes of the (2+1)D interpolating map (see Fig.~\ref{DIII_boundary})  
	
	\subsection{Class AII}
	For class AII with the $\bb{Z}-$indexed parent $d=4$, the converged classification from the AHSS is as follows
	\begin{center}
		\begin{tabular}{ >{\centering}   m{0.12\textwidth}|  >{\centering}   m{0.05\textwidth}   >{\centering}   m{0.05\textwidth} >{\centering}   m{0.05\textwidth} >{\centering\arraybackslash}   m{0.05\textwidth} }
			\hline
			$E_\infty^{p,-(4+p)}$ & $p=0$ & 1 & 2 & 3\\ 
			\hline
			$d=3$ & $\bb{Z}$ & 0 & $\bb{Z}_2^4$ & 0\\
			$d=2$ & $\bb{Z}$ & 0 & $\bb{Z}_2$ & -  \\
			\hline
		\end{tabular}
	\end{center}  
	We do not perform the spectral sequence for $d=4$, but by analogy to the case $d=2$ of class $D$, the $\bb{Z}$ classification is similar to a (4+1)D class A return map. On the other hand, the $\bb{Z}_2$ classifications of the (3+1)D and (2+1)D descendants are encoded in HsPs.    
	
	\subsubsection{Four-dimensional parent}
	Similar to class A, (4+1)D class AII is also classified by the 5D winding number. The time-reversal symmetry enforces $\omega_5(\kk,t)=\omega_5(-\kk,T-t)$ and does not trivialize the integral over the space-time BZ. The dynamic topological invariant is simply
	\begin{equation}
	\kappa_{4D}^{AIII} = \nu_5[R].
	\end{equation} 
	This winding number corresponds to the number of 3D Dirac modes across the quasienergy $\pgap$.    
	
	\subsubsection{Three-dimension descendant}
	\begin{figure*}
		\includegraphics[width=0.8\textwidth]{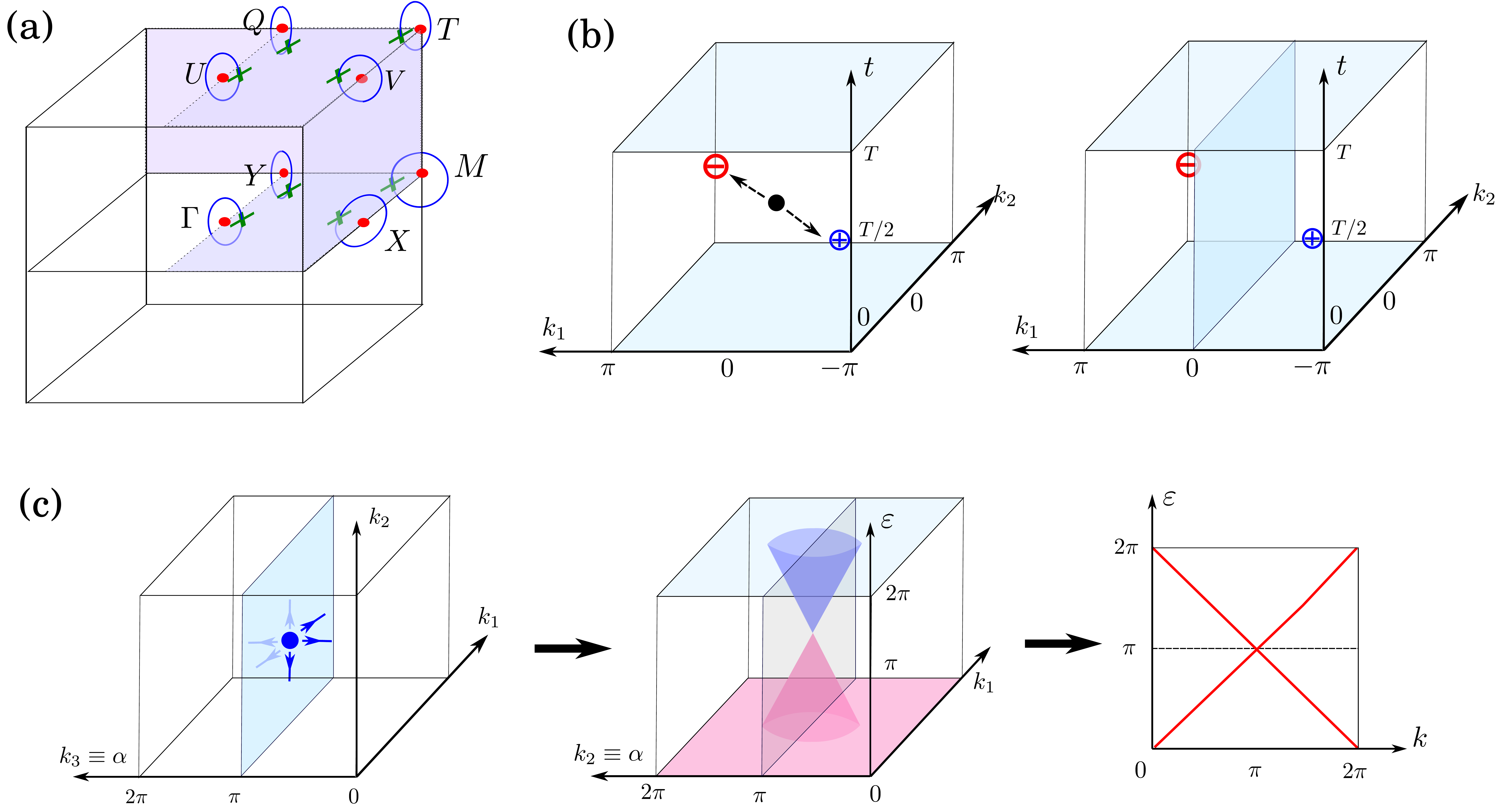}
		\caption{(a) The four-fold degenerate points at HsMs can be generically deformed into two-fold degenerate rings threading through the adjacent HsPs. The ring actually exist in the (3+1)D space-time but the time dimension is not shown for visual clarity. (b) On a (2+1)D BZ, a four-fold singularity at HsM and $t=T/2$ can be split into a pair of two symmetric two-fold singularities with opposite charges. Since the $\pgap$ at $k_1=0$ is open, the return map there can be trivialized so that the 3D winding number over the HsP (half the BZ) is well-defined. (c) Anomalous boundary behaviors of members in the class AII dimensional hierarchy: 3D Dirac mode for the (4+1)D parent, time-reversal-invariant 2D Dirac modes for the (3+1)D descendant, and 1D quantum spin Hall modes for the (2+1)D descendant.\label{AII_diagram}}
	\end{figure*}
	
	We begin the dimensional reduction by constructing the (4+1)D symmetry-preserving interpolating map 
	\begin{equation}\label{3DclassAII}
	\begin{split}
	&\tilde{R}(\kk,0,t) = \mathbbm{1}, \quad \tilde{R}(\kk,\pi,t) = R(\kk,t)\\
	&\theta\tilde{R}(\kk,\alpha,t)\theta^{-1} = \tilde{R}^*(-\kk,2\pi-\alpha,T-t)
	\end{split}
	\end{equation}
	Similar to the class D dimensional hierarchy, the TRS constraints $\nu_5(\tilde{R})$ to have conserved parity, establishing the $\bb{Z}_2$ classification. Without compromising of generality, we assume an explicit form for the four-fold degenerate Dirac mode in (3+1)D class AII phase band
	\begin{equation}
	h(\delta\kk,\delta t) =  \sum_{i=1}^3 \delta k_i \Gamma_i + \delta t\Gamma_4.
	\end{equation}
	The only gap-opening mass terms $\propto \Gamma_5$ are not allowed by the TRS $\theta=\Gamma_{25}$. However, time-reversal-symmetric terms $\propto \Gamma_{i,j}$ with $1\le i<j\le 5$, even though do not open the gap, generically split the degenerate point into a nodal loop inducing $\pgap$ closing points into the adjacent (2+1)D HsPs as schematically shown in Fig.~\ref{AII_diagram}(a). Similar to class DIII hierarchy, this fact suggests that the four-fold singularity charge is translated to invariants defined on the 2-cells, consistent with the AHSS analysis.  
	
	A (2+1)D space-time BZ is naturally characterized by the 3D winding number. We face the same problem that the 2-cell is not periodic, but we note that any $\pgap$-closing points at the (1+1)D boundary can be moved symmetrically to the 2-cell bulk, allowing us to trivialize the boundary and close the 2-cell [see Fig.~\ref{AII_diagram}(b)]. For a 2-cell over $(k_1,k_2,t)$ (we suppress the notation for the perpendicular coordinates) where $k_1\in(0,\pi)$ while $k_2,t\in(-\pi,\pi)$, 
	we define the symmetrized return map over the 2-cell as
	\begin{equation}
	\begin{split}
	\bar{R}(k_1,k_2,t) &= R(k_1,k_2,t)\left[R(\pi,k_2,t)\right]^{-k_1/\pi}\\ &\qquad  \left[R\left(0,k_2,t\right)\right]^{-1+k_1/\pi}.
	\end{split}
	\end{equation}
	As a result, we can define the 3D winding number of the closed 2-cell. From the interpretation sketched in Fig.~\ref{AII_diagram}(a), the classification of the (3+1)D descendant is the parity of the total number of 5D Dirac modes presented in the (4+1)D interpolating map, the bulk-boundary correspondence is thus expressed through
	\begin{equation}\label{3D_AII}
	\kappa_{3D}^{AII} = \nu_3[\bar{R}(\overline{\Gamma Y M X})] + \nu_3[\bar{R}(\overline{U T V Q})] \mod 2
	\end{equation} 
	The invariant $\kappa_{3D}^{AII}= 0$ (1) indicates the trivial (topological) (3+1)D class AII return map. From the dimensional hierarchy, the anomalous boundary hosts 2D time-reversal-symmetric Dirac modes as shown in Fig.~\ref{AII_diagram}(c). The other $\bb{Z}_2$ indices characterize weak phases associated with the stacking of topological 2D slices.
	
	\subsubsection{Two-dimensional descendant}
	The (2+1)D class AII return map can be classified by a symmetry-preserving interpolating map $\tilde{R}(\kk,\alpha,t)$ such that
	\begin{equation}
	\begin{split}
	&\tilde{R}(\kk,0,t) = \mathbbm{1},\quad \tilde{R}(\kk,\pi,t) = R(\kk,t)\\
	&\theta\tilde{R}(k,\alpha,t)\theta^{-1} = \tilde{R}^*(-k,2\pi-\alpha,T-t).
	\end{split}
	\end{equation}   
	The interpolating map realizes a (3+1)D class AII return map and is thus characterized by the $\bb{Z}_2$ index defined in Eq.~\eqref{3D_AII}. The two HsPs appearing in  Eq.~\eqref{3D_AII} correspond to slides at $\alpha=0$ and $\pi$ with the former being trivial by construction. The $\bb{Z}_2$ index, therefore, does not depend explicitly on the interpolating map and is given by
	\begin{equation}
	\kappa_{2D}^{AII} = \nu_3[\bar{R}(\overline{\Gamma Y M X})]\mod 2.
	\end{equation}
	The topological phase with $\kappa_{2D}^{AII}=1$ hosts quantum spin Hall modes at the boundary as a result of the 2D boundary Dirac modes of the interpolating map [see Fig.~\ref{AII_diagram}(c)]. 
	
	In Tab.~\ref{tab2}, we summarize the first-order anomalous Floquet dynamic topological invariant and the boundary signature for all the AZ classes we just study. In the present literature concerning Floquet systems belonging to real AZ classes, we are only aware of the AFT bulk-boundary correspondence for (1+1)D class D \cite{Jiang2011} and (2+1)D class AII \cite{Carpentier2015}. We note that in Ref.~\cite{Carpentier2015}, the topological invariant for (2+1)D class AII is the 3D winding number computed over half the space-time BZ with $t\in(0,T/2)$. Our presented derivation, on the other hand, systematically covers a wide range of system and  explicitly shows the dynamic bulk-boundary correspondence as well as the relation among dimensions.
	
	Before ending the section, we mention the role the static Floquet part of the evolution unitary in the boundary behavior. We first start with the parent system with $\bb{Z}$-indexed AFT characterized by the return map winding number, i.e. AZ class A, and parent systems of classes D and AII. The winding number is naturally related to the Chern number by \cite{Rudner2013}
	\begin{equation}\label{eq34}
	\nu_{2n+1}[R_\varepsilon] - \nu_{2n+1}[R_{\varepsilon'}] = \mathcal{C}_n(\varepsilon\varepsilon'),
	\end{equation} 
	where $\nu_{2n+1}[R_\varepsilon]$ is the winding of the returning map obtained from the unitary $U$ with the branch cut defined at quasienergy $\varepsilon$, and $\mathcal{C}_n(\varepsilon\varepsilon')$ is the $n-$Chern number of the Floquet bands (by diagonalizing $U$) between the quasienergies $\varepsilon$ and $\varepsilon'$. By setting $\varepsilon'=\pi$ and $\varepsilon=0$, it is clear that the number of boundary modes crossing the 0-gap is the number of modes across the $\pgap$ (computed from the AFT index) modified by the Chern number of Floquet bands from 0 to $\pi$. Not surprisingly, Eq.~\eqref{eq34} also has an analog for chiral systems, e.g. class AIII and parent systems of classes BDI and DIII. The winding number is redefined with chiral basis and the Chern number is substituted by the ``chiral winding number" (not to be confused with the phase band winding number)\cite{Fruchart2016}. In short, if the Floquet bands are trivial (zero Chern number or chiral winding number), the $\bb{Z}$ index we define for AFT counts the number of boundary modes at both the $0$ and $\pi-$gaps. Advancing along the dimensional hierarchy, since the boundary of the descendant can be constructed iteratively from the ascendant boundary, the $\bb{Z}_2$ indices for AFT characterizes the simultaneous presence or absence of boundary modes at both the $0$ and $\pi-$gaps. On the other hand, if the Floquet bands are non-trivial, the boundary behaviors at quasienergy gaps are
	\begin{equation}\label{nontrivial_Floquet}
	n(\pi) = \kappa_{\text{AFT}}, \quad n(0) = \kappa_{\text{AFT}} + \kappa_{\text{F}},  
	\end{equation}
	where $n(0,\pi)$ is the number of boundary modes at the $0$ and $\pi-$gaps, $\kappa_{\text{AFT}}$ is the index for AFT obtained from the winding number of the return map, and $\kappa_{\text{F}}$ is the topological index of the static $H_F$ gapped around the 0 quasienergy. The sum is always well defined because  $\kappa_{\text{AFT}}$ and $\kappa_{\text{F}}$ are either both $\bb{Z}$ or both $\bb{Z}_2$ indices.    
	
	\begin{table*}
		\begin{center}
			\begin{tabular}{ >{\centering}   m{0.08\textwidth}|  >{\centering}   m{0.1\textwidth}   >{\centering}   m{0.1\textwidth} >{\centering}   m{0.3\textwidth} >{\centering\arraybackslash}   m{0.3\textwidth} }
				\hline
				AZ class & $d$ & classification & phase-band invariant & boundary signature \\ 
				\hline
				\multirow{1}{0.08\textwidth}{\centering BDI} & 1 & $\bb{Z}$ & $\nu_1[R_{C+}(T/2)]$ & 0D end modes \\
				\hline
				\multirow{2}{0.08\textwidth}{\centering D} & 2 & $\bb{Z}$   & $\nu_3[R]$ & 1D chiral modes \\
				& 1 & $\bb{Z}_2$ & $\nu_1$ of one PH partner $\mod 2$ & 0D majorana end modes \\
				\hline
				\multirow{3}{0.08\textwidth}{\centering DIII} & 3 & $\bb{Z}$ & $\nu_3[R_{C+}(T/2)]$ & 2D CS Dirac modes\\
				& 2 & $\bb{Z}_2$ & $\sum_a \nu_1[\bar{R}_{C+}(a,T/2)] \mod 2$ & 1D helical modes\\
				& 1 & $\bb{Z}_2$ & $\nu_1[\bar{R}_{C+}(\overline{\Gamma X},T/2)] \mod 2$ & 0D majorana Kramers pairs\\
				\hline
				\multirow{3}{0.08\textwidth}{\centering AII} & 4 & $\bb{Z}$ & $\nu_5[R]$ & 3D Dirac modes\\
				& 3 & $\bb{Z}_2$ & $\sum_b \nu_3[R(b)] \mod 2$ & 2D TRS Dirac modes\\
				& 2 & $\bb{Z}_2$ &  $\nu_3[R(\overline{\Gamma Y M X})] \mod 2$ & 1D QSH modes\\
				\hline   			
			\end{tabular}
		\end{center}
		\caption{First-order AFT bulk-boundary correspondence for different AZ classes. The number of spatial dimensions is denoted by $d$. For class DIII, $d=2$, $a=\overline{\Gamma X},\overline{YM}$; while for class AII, $d=3$, $b=\overline{\Gamma Y M X}, \overline{\Gamma Y M X}$. \label{tab2}}    
	\end{table*} 
	
	\section{Second-order AFT protected by inversion symmetry}\label{secondorder}
	In this section, we expand the bulk-boundary correspondence derivation scheme based on the AHSS and dimensional reduction to the second-order AFT protected by the inversion symmetry. For static insulator/superconductors, even when the first-order topology is trivial, the inversion symmetry (or in general spatial symmetry) may obstruct the system to be deformed into the atomic limit, giving rise to higher-order anomalous boundary modes of dimensionality less than $d-1$ ($d$ is the number of bulk spatial dimensions). In this section, we assume the system has an inversion symmetry $\mathcal{I}$ characterized by: $\mathcal{I}R(\kk,t)\mathcal{I}^{-1}=R(-\kk,t)$, $\mathcal{I}^2=\mathbbm{1}$, $[\mathcal{I},\theta\mathcal{K}]=0$, and $\{\mathcal{I},P\mathcal{K}\}=0$.   The second-order topology in a $d-$D space can be viewed as the manifestation of a topological $(d-1)$D subspace. The necessary condition is thus the $(d-1)$D subspace must have non-trivial classification. 
	Therefore, it is interesting that a $d-$D system with trivial first-order classification might have non-trivial second-order AFT, e.g. $d=2$ in class BDI is strictly trivial but the $d=1$ has $\bb{Z}$ classfication. Naively, the second-order topology should inherit the same $\bb{Z}$ classification; however, the $(d-1)$D topological system is actually embedded in a higher-dimensional space so some of the phases can be connected, potentially reducing the classification to $\bb{Z}_n$ \cite{Zhang2020}. We leave this case for future work, focusing only on the second-order topology of the parent and the first descendant in AZ hierarchies D, DIII and AII. Therefore, the second-order AFT, if exist, must have $\bb{Z}_2$ classification,
	
	With gapped Hamiltonians, the higher-order topology is usually studied by the symmetry data at HsMs excluding those associated with the corresponding atomic limit \cite{Po2017,Ono2018,Khalaf2018b,Yu2021}. However, in some cases, the symmetry data at HsMs is trivial and the topological characteristics is actually encoded in higher-dimensional subspaces \cite{Chen2021}. In this work, we do not assume that higher-order topology depends on information at HsMs, but explicitly prove that fact through the AHSS. An important note in computing the $E$-pages is that because $(\mathcal{I}P\mathcal{K})^2=-(P\mathcal{K})^2$, for cells with dimensionality $>0$, classes BDI/D/DIII change to CII/C/CI. With that in mind, we present the converged pages for AZ classes D, DIII and AII with inversion symmetry. The row for class AII and $d=4$ is not derived explicitly but inductively from the $d=3$ case.
	
	\begin{center}
		\begin{tabular}{ >{\centering}   m{0.05\textwidth}|  >{\centering}   m{0.05\textwidth}   >{\centering}   m{0.05\textwidth} >{\centering}   m{0.05\textwidth} >{\centering}   m{0.05\textwidth} >{\centering}   m{0.05\textwidth} >{\centering\arraybackslash}   m{0.05\textwidth} }
			\hline
			class & $d$ & $p=0$ & 1 & 2 & 3 & 4 \\ 
			\hline
			\multirow{2}{0.05\textwidth}{\centering D} & 2 & $\bb{Z}^4$ & 0 & $\bb{Z}$ & - & -\\
			& 1 & $\bb{Z}^2$ & 0 & - & - & -\\
			\hline
			\multirow{3}{0.05\textwidth}{\centering DIII} & 3 & $\bb{Z}^8$ & 0  & 0 & $\bb{Z}$ & - \\
			& 2 & $\bb{Z}^4$ & 0  & 0 & - & -\\
			& 1 & $\bb{Z}^2$ & 0  & - & - & -\\
			\hline
			\multirow{3}{0.05\textwidth}{\centering AII} & 4 & $\bb{Z}^{17}$ & 0  & 0  & 0 & $\bb{Z}$ \\
			& 3 & $\bb{Z}^9$  & 0  & 0  & 0 & - \\
			& 2 & $\bb{Z}^5$  & 0  & 0  & - & - \\
			\hline		
		\end{tabular}
	\end{center}    
	Except for the parent system having an additional $\mathbbm{Z}-$index defined in the bulk, all other topological information is encoded in the HsMs. One can also perform the AHSS on an open geometry, showing that there exist an extra $\mathbbm{Z}_2$-indexed boundary mode corresponding to the second-order topology \cite{Okuma2019}. For class AII, each irrep at a HsM can in principle have an independent $\bb{Z}$ index, resulting in $\bb{Z}^{2^{d+1}}$. However, the first differential restricts $n_{\mathcal{I}+}+n_{\mathcal{I}-}=\text{const}$ for all HsMs, reducing the classification to $\bb{Z}^{2^{d}+1}$. We remind that the AHSS analysis is identical between gapped Hamiltonians and Floquet unitaries within the same AZ class and spatial dimensionality.
	
	\subsection{Class D}
\begin{figure}
		\includegraphics[width=0.48\textwidth]{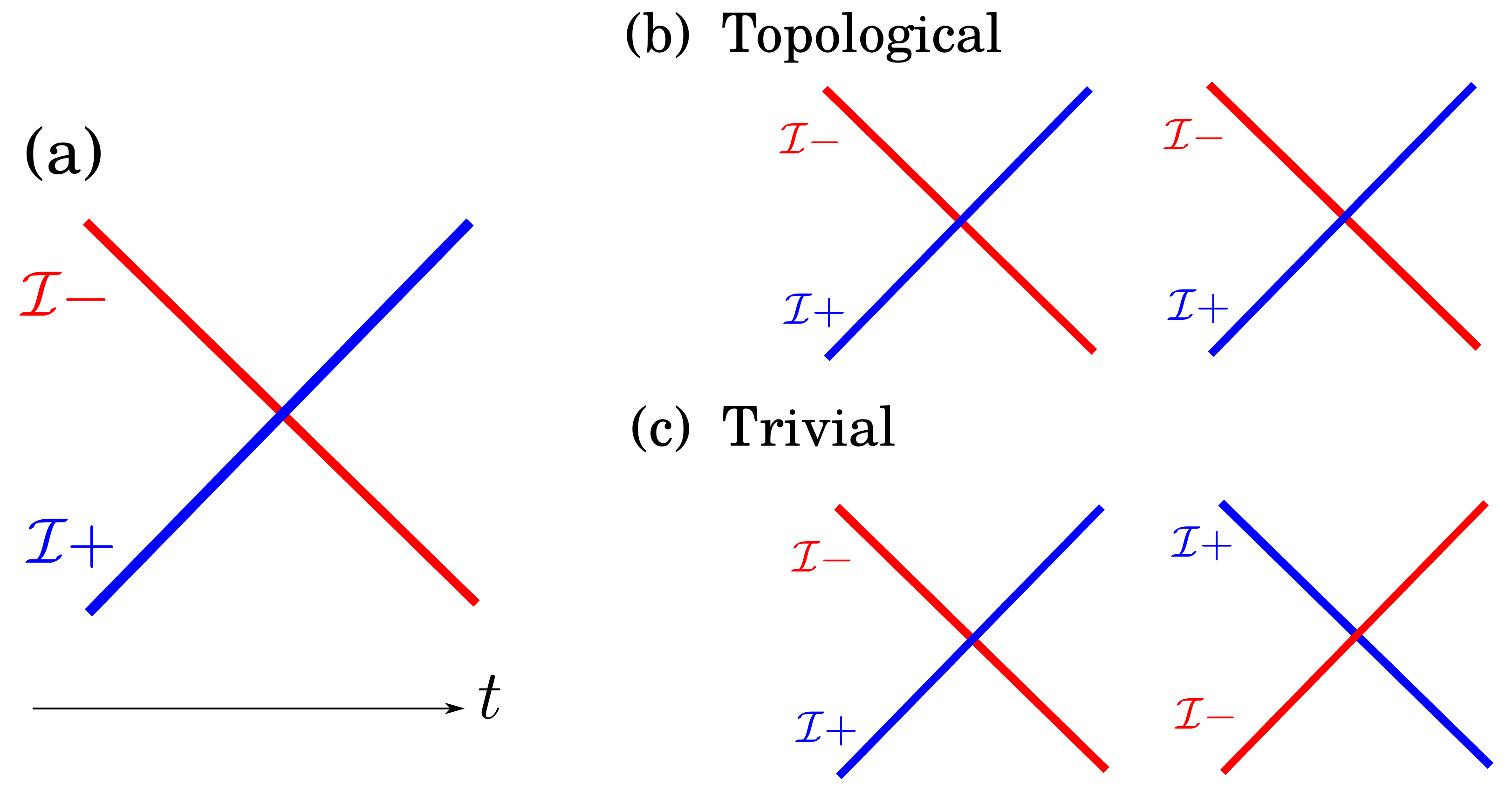}
		\caption{(a) A (2+1)D class D singular point at a HsM projected onto the $t-$axis with the bands labeled by the inversion symmetry eigenvalues. (b) A topological NCP whose gap closing is protected by the inversion symmetry. (c) A trivial NCP whose degeneracy can be lifted by a symmetry-preserving mass term. \label{highorder_sing}}
	\end{figure}

	The inversion symmetry only flips two spatial dimension so the winding number in (2+1) space-time is not trivialized. As a result, the definition of the first-order AFT bulk $\bb{Z}$ index is unchanged. The classification of the second-order AFT is only meaningful if the first-order one is trivial, corresponding to the vanishing 3D winding number $\nu_3[R]$. This situation can be interpreted in two ways: (i) the phase band has no $\pgap$ singular points or (ii) the $\pgap$ singular points exist in null-charge pairs (NCP), i.e. a pair of two singularities with opposite charges. Without any spatial symmetry, (i) and (ii) are adiabatically connected, but this is no longer true if the inversion symmetry (or other appropriate spatial symmetries) is presented. We first note that off the HsM axes, singularities must exist in pairs by the PHS, and two pairs with opposite charge can always annihilate each other as the inversion symmetry does not manifest off-axis. On the other hand, at the HsM axes, a NCP is robust if it is described by
	\begin{equation}\label{topo_point}
	h(\delta \kk,\delta t) = \delta k_1\Gamma_4-\delta k_2\Gamma_{34} + \delta t\Gamma_{45},
	\end{equation}
	in the basis where $\mathcal{I}=\Gamma_{45}$ and $P=\Gamma_4$. Due to anti-commutation between the inversion and PH symmetries, PH-related bands have opposite inversion eigenvalues [Fig.~\ref{highorder_sing}(a)].  Since $\mathcal{I}\propto h(0,\delta t) $, any mass term anti-commuting with $h$ is thus forbidden by the inversion symmetry. On the other hand, within the same basis, another NCP written as 
	\begin{equation}\label{trivial_point}
	h(\delta \kk,\delta t) = \delta k_1\Gamma_4-\delta k_2\Gamma_{5} + \delta t\Gamma_{5}.
	\end{equation}   
	can be gapped out by a mass term $\propto \Gamma_{42}$.  The difference between Eqs.~\eqref{topo_point} and \eqref{trivial_point} is apparent in the projection onto the $t-$axis. Specifically, the former has the 1D winding number of the $\mathcal{I}+$ subspace $\nu_{1}[R_{\mathcal{I}+}(\text{HsM})]=2$, while the latter has $\nu_{1}[R_{\mathcal{I}+}(\text{HsM})]=0$ [see Fig.~\ref{highorder_sing}(b) and (c)]. Naively, the description~\eqref{topo_point} gives rise to a $\bb{Z}$ classification because if multiple of its copies are stacked together, the gap closing is still robust by the similar argument. This is indeed reflected in the $\bb{Z}$ classification at HsMs produced by the AHSS. However, following the argument for symmetry indicators, the atomic limit needs to be subtracted from these $\bb{Z}$ indices, resulting in the actual $\bb{Z}_2$ classification of the higher-order topological phase \cite{Po2017,Ono2018,Khalaf2018b,Yu2021}. 
	
	In this work, we arrive at the $\bb{Z}_2$ classification not by modding out the atomic limit but by introducing a spatially modulated $\pgap$-opening term, effectively bringing out from the (2+1)D bulk a topological (1+1)D subsystem whose $\bb{Z}_2$ classification is shown in the previous section. This process can be extended naturally along the dimensional hierarchy, providing a fast way to derive the dynamic bulk-boundary correspondence. In the case described in Eq.~\eqref{topo_point}, the mass term is given by $m(x^1) = M\text{sign}(x^1)\Gamma_{42}$, preserving the inversion symmetry globally as $\mathcal{I}m(x^1)\mathcal{I}^{-1}=m(-x^1)$ (here $x^1$ conjugates to $k_1$). At the edge $x^1=0$, modes crossing the branch cut are obtained by solving the equation
	\begin{equation}\label{edgemode}
	[-\partial_{1}+M\Gamma_{2}\text{sign}(x^1)]\ket{\psi} = 0.
	\end{equation}
	Without loss of generality, we assume $M>0$ so Eq.~\eqref{edgemode} has two solutions $\ket{\psi_{1,2}}e^{-|x^1|/M}$ with $\ket{\psi_{1,2}}$ being two eigenvectors of $\Gamma_2$ with eigenvalues 1. Projecting \eqref{topo_point} onto two $\pgap$ modes gives the effective edge as
	\begin{equation}\label{domain_wall}
	h(\delta k_2,t) = \delta k_2 \sigma_1 +\delta t \sigma_3
	\end{equation}
	with $P=\sigma_1$. This describes a topological (1+1)D class D return map with $\bb{Z}_2$ classification indexed by the parity of the total number of $\pgap$ closing points. Thus, the second-order AFT index is given by
	\begin{equation}\label{eq_etaD}
	\eta^{D}_{2D} = \frac{1}{2}\sum_{\text{HsM}} \nu_1 [R_{\mathcal{I}+}(\text{HsM})] \mod 2.
	\end{equation}
	The factor $1/2$ is founded on our dimensional reduction argument that two singularities of opposite charges produces one symmetry-protected singularity on the domain wall [the $4\times4$ gapless matrix (Eq.~\ref{topo_point}) reduces to the $2\times2$ gapless matrix on the domain wall (Eq.~\ref{domain_wall}).] Because of the factor, the index is only well-defined when $\sum \nu_1 [R_{\mathcal{I}+}(\text{HsM})]\equiv0 \mod 2$, otherwise $\sum \nu_1[R_{\mathcal{I}+}(\text{HsM})]\equiv1 \mod 2$ corresponds to the non-zero 3D winding number, indicating the first-order AFT.
	
	The inversion symmetry, even thought does not change the classification of the (1+1)D class D return map, provides an additional expression for its indicator. Without the any spatial symmetries, the topological invariant is the 1D winding number of one PH partner defined from the continuity of the phase band. When the inversion symmetry is introduced, two bands related by PHS must have opposite inversion eigenvalue, simplifying the first-order AFT indicator to
	\begin{equation}
	\eta^{D}_{1D} = \sum_{\text{HsM}} \nu_1 [R_{\mathcal{I}+}(\text{HsM}))] \mod 2,
	\end{equation} 
	with the bulk-boundary correspondence identical to the $\kappa_{1D}^D$ without the inversion symmetry.
	
	\subsection{Class DIII}
	\subsubsection{$\bb{Z}_2$ index for the (3+1) parent}
	As discussed earlier, it is more natural to think of class DIII singularities as existing in an enlarged space with the extra dimension being flipped by the chiral symmetry. We choose the action of the inversion symmetry so that
	\begin{equation}
	\mathcal{I}R(\kk,t,\beta)\mathcal{I}^{-1}=R(-\kk,t,-\beta)
	\end{equation}
	A 5D singularity in this enlarged space is robust by the fact that $\mathcal{I}$ flips four momentum and does not trivialize $\nu_5$. Similar to case of class D, a NCP at a HsM can be protected by the inversion symmetry, giving rise to the symmetry-protected second-order topology. 
	
	Because $\{P\mathcal{K},\mathcal{I}\}=0$ and $[\theta\mathcal{K},\mathcal{I}]=0$, along $\kk = \text{HsM}$, and $\beta=0$, the two right-moving bands must have the same inversion eigenvalue, opposite to that of the other two left-moving bands. Schematically, the projection of a singularity at HsM on the $t-$axis is also described by Fig.~\ref{highorder_sing}(a) except for that each band is now a two-component Kramers pair by the TRS. Therefore, $h(0,\delta t,0)\propto\delta t \mathcal{I}_4$ where the subscript 4 denotes the projection of $\mathcal{I}$ onto the 4-band subspace. If two singularities have opposite charges but their projections on the $t-$axis are similar, i.e. the NCP is written as
	\begin{equation*}
	h_{\text{comp}}(0,\delta t, 0) \propto \delta t  \sigma_0\otimes \mathcal{I}_4 = \delta t \mathcal{I}_8,
	\end{equation*}
	any mass terms must anti-commute with the inversion symmetry operator, and is thus not allowed. We now show that this inversion-protected $\pgap$ closing leads to a topological (2+1)D subsystem. For demonstration, we explicitly construct a NCP pinned at $(\text{HsM},T/2,0)$ as
	\begin{equation}\label{DIII_nodalpair}
	\begin{split}
	h_{\text{comp}}&(\delta\kk,\delta t,\beta) = \sigma_3\Gamma_1\delta k_1  \\
	&+\sigma_0\left(\sum_{i=2,3}\delta k_i\Gamma_i + \delta t\Gamma_4+\beta\Gamma_5\right).
	\end{split}
	\end{equation} 
	with $C=\sigma_0\Gamma_{45}$, $P=\sigma_0\Gamma_{24}$, and $\mathcal{I}=\sigma_0\Gamma_{4}$. A homogeneous mass term $\propto \sigma_1\Gamma_1$ anti-commutes with $\mathcal{I}$ and is thus forbidden. However, a symmetry-preserving spatially modulated mass term can be constructed accordingly as $m(x^1)=M\text{sgn}(x^1)\sigma_1\Gamma_1$. Similar to the previous case of class D, this mass term generate a gapless domain wall where the effective Hamiltonian can be obtained by projecting the original instantaneous Hamiltonian onto $\pgap$-crossing modes which are the four eigenvectors with eigenvalues 1 of $\sigma_1\Gamma_1$, resulting in
	\begin{equation}\label{DIII_gapclosing point}
	h(\delta k_{2,3},\delta t, \beta) = \sum_{i=2,3}\delta k_i\Gamma_i + \delta t\Gamma_4+\beta\Gamma_5,
	\end{equation}
	with $C=\Gamma_{45}$ and $P=\Gamma_{24}$. This describes the return map of the topological (2+1)D class DIII, leading to the second-order topological bulk-boundary correspondence
	\begin{equation}\label{DIII_nu2}
	\eta_{3D}^{DIII} =  \frac{1}{4}\sum_{\text{HsM}}  \nu_1 [R_{\mathcal{I}+}(\text{HsM})] \mod 2,
	\end{equation}
	where compared with Eq.~\eqref{eq_etaD} the extra $1/2$ factor accounts for the Kramers degeneracy, and $\eta_{3D}^{DIII}=1(0)$ indicates the presence (absence) of 1D anomalous boundary modes in a 3D open geometry. 
	
	\subsubsection{$\bb{Z}_4$ index for (2+1)D descendant}
	
	A (3+1)D class DIII return map with second-order topology, under the dimensional reduction, produces a (2+1)D return map hosting a topological (1+1)D subspace. The process is performed similarly to other presented cases with one momentum of the (3+1)D return map being used as the interpolating parameter $\alpha$ so that the second-order AFT index of the interpolating map $\bar{R}(\kk,\alpha,t)$ is
	\begin{equation}
	\eta_{3D}^{DIII} = \frac{1}{4} \sum_{\alpha=0,\pi}\sum_{\text{HsM}} \nu_1[\bar{R}_{\mathcal{I}+}(\alpha,\text{HsM})] \mod 2.
	\end{equation}
	By construction, $\bar{R}(0,\kk,t)=\mathbbm{1}$ and $\bar{R}(\pi,\kk,t)=R(\kk,t)$ so for the (2+1)D class DIII return map, the second-order AFT index is defined identically to Eq.~\eqref{DIII_nu2} except for the sum is taken over 2D high-symmetry momenta.
	
	There exist, however, an important difference from the parent, i.e. the topological invariant characterizing the first-order AFT cannot be defined by Eq.~\eqref{2DclassDIII} because the HsLs are all trivial as shown by the AHSS. In the (2+1)D class DIII return map, the $\pgap$ singularities that give rise to the first-order AFT generically exist in the form of nodal rings. Consulting Eq.~\eqref{DIII_singularity}, the nodal ring is nothing but a singular point at a HsM deformed by a perturbation $\propto \Gamma_{45}$. If the inversion symmetry is presented, in this case $\mathcal{I}=\Gamma_5$, it rules out such perturbations and stabilizes the singular point. As such, the natures of the first-order AFT in the (2+1)D descendant with and without inversion-symmetry are indeed identical.
The bulk-boundary correspondences for the first and second-order AFT are now both encoded in the HsMs and can be obtained straightforwardly from the (3+1)D parent through the dimensional hierarchy
		\begin{equation*}
		\frac{1}{2} \sum_{\text{HsM}} \nu_1[R_{\mathcal{I}+}(\text{HsM})] = \begin{cases}
		2n + 1 & \text{ 1\textsuperscript{st}-order topo.}\\
		4n + 2 & \text{ 2\textsuperscript{nd}-order  topo.}\\
		4n     & \text{ trivial topo.}
		\end{cases}.
		\end{equation*} 
	As a result, we can define a $\bb{Z}_4$ index to capture the two phenomena
	\begin{equation}
	\eta_{2D}^{DIII} = \frac{1}{2} \sum_{\text{HsM}}  \nu_1 [R_{\mathcal{I}+}(\text{HsM})] \mod 4.
	\end{equation}
	The dynamic bulk-boundary correspondence is as follows: for $\eta_{2D}^{DIII}=1,3$ the system has first-order AFT and hosts 1D helical modes across the $\pi$-quasienergy; for $\eta_{2D}^{DIII}=2$, the system has second-order AFT with two inversion-symmetric majorana Kramers pairs as the boundary of the (1+1)D subsystem; lastly, for $\eta_{2D}^{DIII}=0$, the boundary is trivial with opened $\pgap$. 
	
	For the (1+1)D descendant, the second-order AFT is not supported; while for the same reason as the (2+1)D case, the indicator for the first-order AFT is modified as
	\begin{equation}
	\eta_{1D}^{DIII} = \frac{1}{2} \sum_{\text{HsM}}  \nu_1 [R_{\mathcal{I}+}(\text{HsM})] \mod 2,
	\end{equation}
	with the value of 1 (0) corresponds to the absence (presence) of the $\pgap$ majorana Kramers pairs.
	
	\subsection{Class AII} 
	\begin{table}
		\begin{center}
			\begin{tabular}{ >{\centering}   m{0.05\textwidth}|  >{\centering}   m{0.05\textwidth} >{\centering}   m{0.1\textwidth} >{\centering}   m{0.1\textwidth} >{\centering\arraybackslash}   m{0.1\textwidth} }
				\hline
				class & $d$ & Classification & 1$^\text{st}-$order & 2$^\text{nd}-$order  \\ 
				\hline
				\multirow{2}{0.05\textwidth}{\centering D} & 2 & $\bb{Z}\times\bb{Z}_2$ & $\checkmark$ & $\checkmark$ \\
				& 1 & $\bb{Z}_2$ & $\checkmark$ & -\\
				\hline
				\multirow{2}{0.05\textwidth}{\centering DIII} & 3 & $\bb{Z}\times \bb{Z}_2$ & $\checkmark$ & $\checkmark$\\
				& 2 & $\bb{Z}_4$ & $\checkmark$ & $\checkmark$\\
				& 1 & $\bb{Z}_2$ & $\checkmark$ & -\\
				\hline
				\multirow{2}{0.05\textwidth}{\centering AII} & 4 & $\bb{Z}\times \bb{Z}_2$& $\checkmark$ & $\checkmark$\\
				& 3 & $\bb{Z}_4$ & $\checkmark$& $\checkmark$\\
				& 2 & $\bb{Z}_2$ & $\checkmark$& -\\
				\hline		
			\end{tabular}
		\end{center}
		\caption{Classification of return map with inversion symmetry. The $\bb{Z}$ indices are defined similar to the case without the inversion symmetry while the $\bb{Z}_4$ and $\bb{Z}_2$ indices are defined from the 1D winding number of the $\mathcal{I}$ along the time dimension at HsMs. \label{tab3}} 
	\end{table}

	For the (4+1)D parent system, the inversion symmetry does not trivialize the $\bb{Z}$ index of the first-order AFT as it flips four momenta, thus keeping the sign of the winding number density. A Dirac mode in the (4+1)D space-time can be written as
	\begin{equation}\label{AII_singularity}
	h(\delta k,\delta t) = \sum_{i=1}^4 \delta k_i\Gamma_i + \delta t\Gamma_5.
	\end{equation}
	The only choices for the time-reversal and inversion symmetries are $\theta=\Gamma_{25}$ and $\mathcal{I} = \Gamma_5$. The form of the inversion symmetry constraint that on the $t$-axis, the two left-moving bands have the same inversion eigenvalue, opposite to those of the two right-moving bands, similar to the case of class DIII. To demonstrate the emergence of a topological (3+1)D subsystem, we can introduce a spatially modulated mass term similar to class DIII. The conclusion is the same: a NCP at HsMs is protected by the inversion symmetry if along the $t-$axis, the winding number of the $\mathcal{I}+$(or $\mathcal{I}-$) is $8n+4$. There is a subtlety as the 4-fold singular point \eqref{AII_singularity} can be generically deform into a nodal ring (or sphere) by a symmetry-preserving perturbation, possibly opening the $\pi$-gap along the $t-$axis. However, all the perturbations must commute with the inversion symmetry operator, or $h(0,\delta t)$, ensuring that the symmetry-preserving nodal rings must cut through the $t-$axis. This establishes that the winding number of  $\mathcal{I}+$ irrep along the time dimension at HsMs is indeed invariant. From this fact, we can straight forwardly defined the second-order AFT $\bb{Z}_2$ index for the (4+1)D parent and the $\bb{Z}_4$ index for the (3+1) descendant. Specifically, for the (4+1)D parent,
	\begin{equation}
	\eta_{4D}^{AII} = \frac{1}{4} \sum_{\text{HsM}} \nu_1[R_{\mathcal{I+}}(\text{HsM})] \mod 2,
	\end{equation} 
	where $\eta_{4D}^{AII}=1$ (0) corresponds to the presence (absence) of the second-order time-reversal-invariant 2D Dirac mode across the quasienergy BZ and is only well defined when $\nu_5[R]=0$ ; while for the (3+1)D descendant,
	\begin{equation}
	\eta_{3D}^{AII} = \frac{1}{2} \sum_{\text{HsM}} \nu_1[R_{\mathcal{I+}}(\text{HsM})] \mod 4,
	\end{equation}
	where $\eta_{3D}^{AII} = 0/1,3/2$  corresponds to the trivial/first-order/second-order AFT respectively. Lastly, for the (2+1)D descendant, the second-order AFT is trivial because the (1+1)D class AII return map is trivial but the indicator for the first-order AFT is nevertheless modified as
	\begin{equation}
	\eta_{2D}^{AII} = \frac{1}{2} \sum_{\text{HsM}} \nu_1[R_{\mathcal{I+}}(\text{HsM})] \mod 2.
	\end{equation}
	Before summing up the section, we clarify the ambiguity in the choice of $\mathcal{I}+$ versus $\mathcal{I}-$ subspace. From the AHSS analysis, $\sum \nu_1[R_{\mathcal{I+}}(\text{HsM})] + \sum \nu_1[R_{\mathcal{I-}}(\text{HsM})] = 2\times 2^d\times n$, where the first factor of 2 is due to the Kramers pairs, $2^d$ is the number of HsMs and $n$ is an integer. As a result, the substitution of $\mathcal{I}+$ by $\mathcal{I}-$ does not change the value of our defined invariants. We summarize the classification for class D, DIII and AII dimensional hierarchies in Tab.~\ref{tab3} where except from the $\bb{Z}$ indices, the other $\bb{Z}_2$ and $\bb{Z}_4$ indices are all derived from the winding number of the $\mathcal{I}+$ subspace along the $t-$axis at HsMs, in the same spirit as the symmetry indicators characterizing gapped Hamiltonians. Lastly, we note that the topology of the Floquet bands can be added to the AFT indicators $\eta$ to determine the boundary behavior across the $0$-gap in the same manner as Eq.~\ref{nontrivial_Floquet}.

	\section{Class DIII demonstrative model}\label{example}
	In this section, we demonstrate our theory for (2+1)D class DIII return maps where the first-order topology is characterized by helical modes traversing the $\pgap$ while the second-order one hosts inversion-symmetric corner majorana Kramers pairs.  Experimentally, class DIII driven systems can be realized on an optical lattice of BEC \cite{Yang2018} or a Josephson junction comprised of a non-magnetic semiconductor sandwiched between two 2D superconductors \cite{Zhang2021}. Here, we do not attempt to propose a realistic model but only provide a minimal tight-binding model that exhibits various AFT phenomena.
	
	\begin{figure*}
		\includegraphics[width=0.98\textwidth]{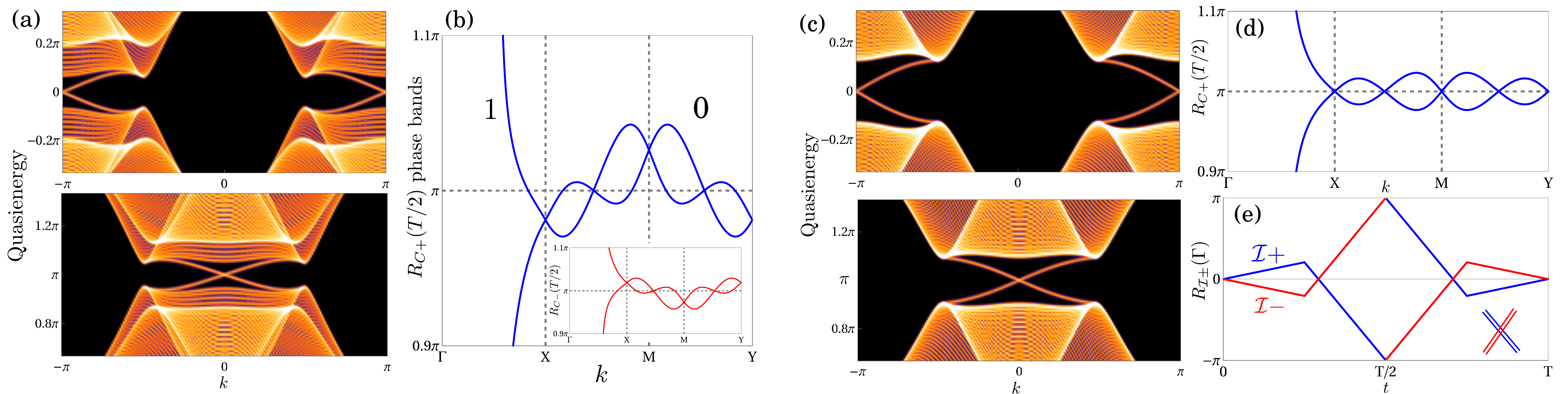}
		\caption{(a) Density of states at the upper edge of a ribbon configuration under a inversion-symmetry-breaking driving Hamiltonian ($m\neq 0$), focusing on the helical AFT modes traversing the zero and $\pi$ quasienergies. (b) Phase bands of $R_{C+}(T/2)$ along high-symmetry lines $\Gamma X$, $XM$ and $MY$. The numbers denote the number of $\pgap$ crossings modulo 2. The inset shows the phase bands of $R_{C-}(T/2)$. (c) Same as (a) but for the inversion-symmetric case $(m=0)$. (d) Symmetry-preserving counterpart of (c) with the 1D winding number of the HsLs identically trivial. (e) Phase bands along the $t-$axis the $M$ momentum with the inset zooming in the four-fold degenerate point at the $\pi$ quasienergy.\label{first_order}}
	\end{figure*}
	
	\subsection{First-order AFT}
	We first demonstrate the first-order AFT with and without the inversion symmetry by the tight-binding model
	\begin{equation}\label{helical_model}
	\begin{split}
	H(\kk,t) &= \left[J(t)(2-\cos k_x -\cos k_y) -\mu\right] s_0\sigma_3\\
	&+\Delta(\sin k_x s_3\sigma_1+\sin k_y s_0\sigma_2) + ms_2\sigma_2 	 
	\end{split}
	\end{equation}
	The Pauli matrices $s$ and $\sigma$ denote the spin and the the particle-hole degrees of freedom so that PHS and TRS are expressed by $P=s_0\sigma_1$ and $\theta=is_2\sigma_0$. Important, the presence of the inversion symmetry $\mathcal{I}=s_0\sigma_3$ ($\{P\mathcal{K},I\}=0$) is controlled by the parameter $m$ such that the Hamiltonian is inversion-symmetric for $m=0$. The Hamiltonian includes common terms: the hopping term $J$, the chemical potential $\mu$ and the odd pairing $\Delta$. The time-dependence is included in the hopping strength by
	\begin{equation}
	J(t) = \begin{cases}
	J_0 - J_D & \text{ for } t\in[0,T/4]\cup[3T/4,T]\\
	J_0 + J_D & \text{ for } t\in(T/4,3T/4)
	\end{cases}
	\end{equation}
	We choose $J_0=2$, $J_D=2\sqrt{2}$, $\mu=-2$, $\Delta = 1$, $T=\pi/4$, and $m=0 (0.5)$ in the cases with(without) inversion symmetry.
	
	In Figs.~\ref{first_order}(a) and (b), we present the simulation results in the inversion-symmetry-breaking case ($m=0.5$). The density of states at the upper edge of a ribbon configuration clearly shows the presence of helical boundary modes at both the 0 and $\pi$ quasienergies, establishing the non-trivial first-order AFT in the (2+1)D class DIII return map. According to our theory, this phase is characterized by the 1D winding number of the $C+$ subspace over the HsLs $\overline{\Gamma X}$ and $\overline{M Y}$. Figure.~\ref{first_order}(b) shows that $\nu_1[\bar{R}_{C+}(\Gamma X,T/2)]=1$ and $\nu_1[\bar{R}_{C+}(MY,T/2)]=0$, resulting in $\kappa_{2D}^{DIII}=1$ consistence with the simulation on the ribbon geometry. If we choose the $C-$ subspace instead, the topological invariant is unchanged up to modulo 2, showing our bulk-boundary correspondence is uniquely defined.
	
	By tuning $m=0$, we impose the inversion symmetry upon the system. Figure.~\ref{first_order}(c) shows that this action does not affect the manifestation of the first-order AFT. However, the same invariant is no longer applicable. As shown in Fig.~\ref{first_order}(d), every time a band cross the $\pi$-gap in one direction, there exists another band crossing the $\pi$-gap in the opposite direction, making the total number of crossings identically zero.  The winding numbers over the HsLs are thus trivialized. This agrees with our analysis that the topological information of the inversion-symmetry-preserving (2+1)D return map is instead encoded in the winding number along the time dimension at fixed HsMs. In Fig.~\ref{first_order}(e), we demonstrate that $\nu_1 [R_{\mathcal{I}+}(M)]=-2$, our phase band analysis also shows $\nu_1[R_{\mathcal{I}+}(X)]=\nu_1[R_{\mathcal{I}+}(Y)]=-2$, establishing $\eta_{2D}^{DIII} = 1$ consistent with the first-order AFT displayed in the open-boundary simulation.
	
	\subsection{Second-order AFT}
	\begin{figure}
		\includegraphics[width=0.48\textwidth]{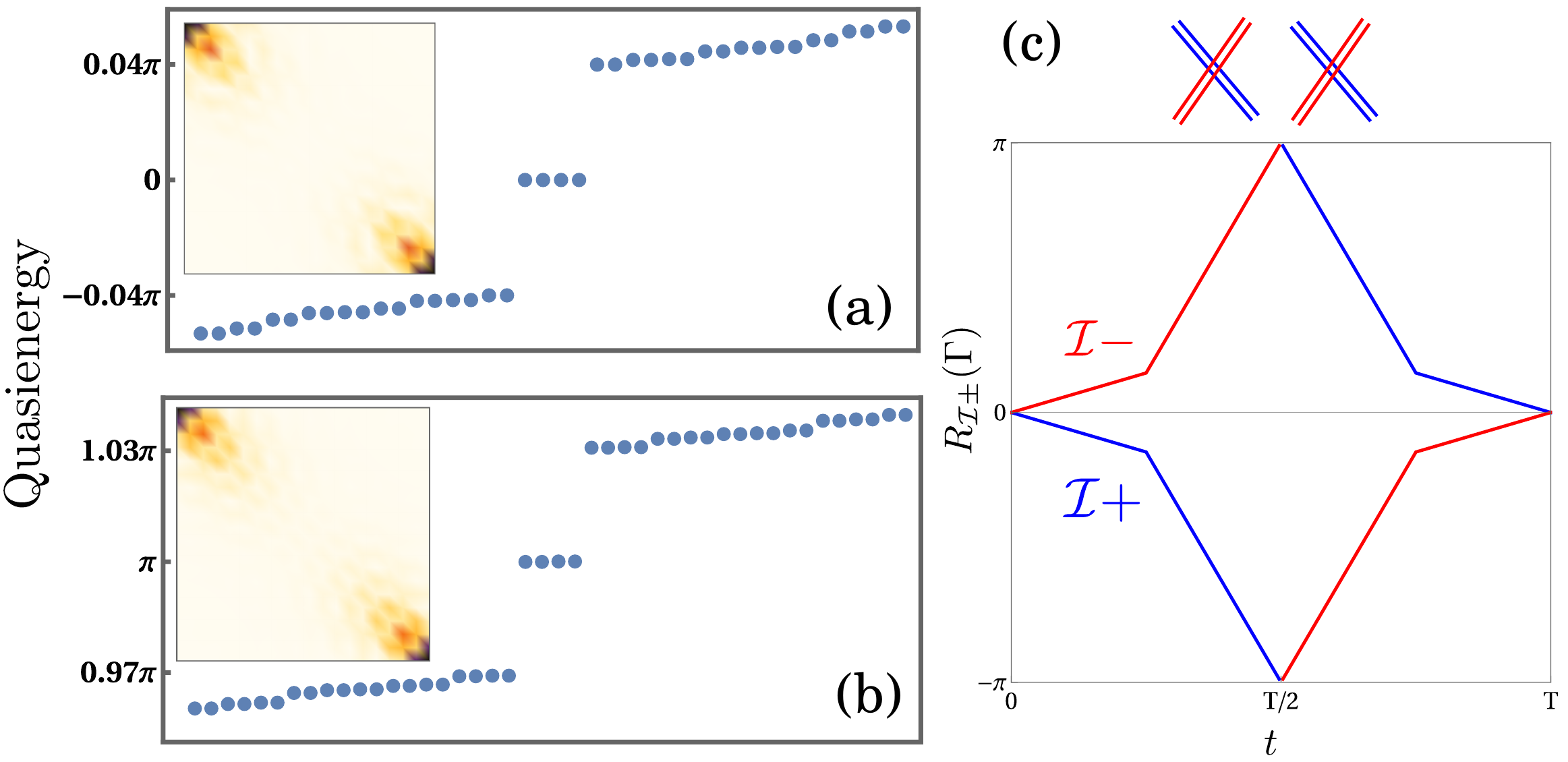}
		\caption{(a,b) Quasienergy levels and the density profile of the in-gap modes at the zero (a) and $\pi$ quasi-energy gaps. (c) Phase bands along the $t-$axis the $M$ momentum, the top subfigure shows the $\pgap$ crossing is actually composed of 8 bands. \label{DIII_corner}}
	\end{figure}   
	
	According to our theory, to realize the second-order AFT, we need at least eight bands. For this reason, we add another orbital degrees of freedom, denoted by the Pauli matrices $\rho$. The driving Hamiltonian is modified as
	\begin{equation}
	\begin{split}
	H(\kk,t) &= \left[J(t)(2-\cos k_x -\cos k_y) -\mu\right] \rho_0s_0\sigma_3\\
	&+\Delta(\sin k_x \rho_3s_3\sigma_1+\sin k_y \rho_0s_0\sigma_2) \\
	& + v(\rho_2s_3\sigma_0+\rho_1 s_0\sigma_3)	 
	\end{split}
	\end{equation}
	The temporal modulation and the numerical value of $J(t), \mu, \Delta \text{ and }T$ are similar to Eq.~\eqref{helical_model}; the last term $v=0.2$ couples two pairs of helical modes, creating the Karmmers pairs of majorna corners. 
	
	Figures.~\ref{DIII_corner}(a) and (b) exhibit the simultaneous existence of majorana Kramers pairs at both the 0 and $\pi$ quasienergies, establishing this phase as the second-order AFT phase. The phase band analysis at HsMs shows that $\nu_1[R_{\mathcal{I}+}(X)]=\nu_1[R_{\mathcal{I}+}(Y)]=\nu_1[R_{\mathcal{I}+}(M)]=-4$ (see Fig.~\ref{DIII_corner} for the phase band at $M$), leading to  $\eta_{2D}^{DIII} = 2$ consistent with our derived bulk-boundary correspondence. Our example has demonstrated the merit of our classification scheme, and we expect our theory to be readily applied into realistic Floquet systems.
	
	\section{Conclusion}\label{conclusion}
	We have derived systematically the dynamic bulk-boundary correspondence along the dimensional hierarchies of classes BDI, D, DIII and AII including both the first-order and second-order AFT protected by the inversion symmetry. We rely on the fact that AFT arises from irremovable branch-cut crossing and study the robustness of these crossings to obtain the classification. Our derivation is organized into sequences of decreasing dimensionality within each AZ class where the parent (highest dimensionality) has $\bb{Z}$ classification related to the total charge of bulk phase band singularities. Advancing along the hierarchy, the descendant system is classified by an interpolating map between it and the identity operator. This map is nothing but the higher-dimensional ascendant return map, allowing us to iteratively derive the classification and the anomalous boundary behavior. Physically, the process can be thought of the passing of phase band singularities from the parent to the descendants. However, the topological invariants characterizing the singularities depend highly on the dimensionality and must be modified accordingly to describe the descendant systems. This is a non-trivial task that so far has only been achieved in isolated occasions. 
	
	Our work avoids the trial-and-error approach by implementing the AHSS which is conventionally used for computing K-group of gapped Hamiltonians. Interestingly, the process works equally well for unitary return map, as a result from the similarity between the ten-fold periodic tables for the gapped Hamiltonian and unitary loops. The AHSS identifies subspace of the BZ where a topological invariant is robust. Specifically, the topological information is encoded in HsMs for (1+1)D class D, in HsLs for (2+1)D and (1+1)D class DIII, and in HsPs for (3+1)D and (2+1)D class AII return maps. By introducing symmetry-preserving perturbation to the phase band singular points, we prove this analysis indeed reflects the physical picture and accordingly derive the dynamic bulk-boundary correspondence.  
	
	With the inversion symmetry, the AFT landscape is greatly enriched with the manifestation of higher-order anomalous boundary modes. In our classification scheme, we regard the second-order AFT as an emergence of a topological lower-dimensional $(d-1)$D subsystem inside the bulk of the $d-$D system. By introducing a spatially modulated $\pgap$ opening term so that it preserves symmetry globally, we explicitly bring out the topological subsystem from the bulk and relate its physical origin to a NCP that is protected by the inversion symmetry. Surprisingly, by studying symmetry-preserving perturbations, we show that these singularities that give rise to the second-order AFT (as well as first-order AFT in descendant systems) are pinned to the time axis at HsMs, consistent with the AHSS analysis. This allows us to obtain the dynamic version of symmetry indicators for the AFT. It is noteworthy that not all systems with a spatial symmetry admit the HsM indicators but our established framework can be adapted in these situations as the AHSS rigorously identifies the topological invariants rather than assuming their existence.
	
	\begin{acknowledgements}
		We are grateful to Sankar Das Sarma and Jiabin Yu for bringing up the problem and helpful discussions. The work is supported by the Laboratory for Physical Science.
	\end{acknowledgements}
	
	\bibliographystyle{apsrev4-1}
	\bibliography{Floquet_classification}   

\begin{thebibliography}{57}%
\makeatletter
\providecommand \@ifxundefined [1]{%
 \@ifx{#1\undefined}
}%
\providecommand \@ifnum [1]{%
 \ifnum #1\expandafter \@firstoftwo
 \else \expandafter \@secondoftwo
 \fi
}%
\providecommand \@ifx [1]{%
 \ifx #1\expandafter \@firstoftwo
 \else \expandafter \@secondoftwo
 \fi
}%
\providecommand \natexlab [1]{#1}%
\providecommand \enquote  [1]{``#1''}%
\providecommand \bibnamefont  [1]{#1}%
\providecommand \bibfnamefont [1]{#1}%
\providecommand \citenamefont [1]{#1}%
\providecommand \href@noop [0]{\@secondoftwo}%
\providecommand \href [0]{\begingroup \@sanitize@url \@href}%
\providecommand \@href[1]{\@@startlink{#1}\@@href}%
\providecommand \@@href[1]{\endgroup#1\@@endlink}%
\providecommand \@sanitize@url [0]{\catcode `\\12\catcode `\$12\catcode
  `\&12\catcode `\#12\catcode `\^12\catcode `\_12\catcode `\%12\relax}%
\providecommand \@@startlink[1]{}%
\providecommand \@@endlink[0]{}%
\providecommand \url  [0]{\begingroup\@sanitize@url \@url }%
\providecommand \@url [1]{\endgroup\@href {#1}{\urlprefix }}%
\providecommand \urlprefix  [0]{URL }%
\providecommand \Eprint [0]{\href }%
\providecommand \doibase [0]{http://dx.doi.org/}%
\providecommand \selectlanguage [0]{\@gobble}%
\providecommand \bibinfo  [0]{\@secondoftwo}%
\providecommand \bibfield  [0]{\@secondoftwo}%
\providecommand \translation [1]{[#1]}%
\providecommand \BibitemOpen [0]{}%
\providecommand \bibitemStop [0]{}%
\providecommand \bibitemNoStop [0]{.\EOS\space}%
\providecommand \EOS [0]{\spacefactor3000\relax}%
\providecommand \BibitemShut  [1]{\csname bibitem#1\endcsname}%
\let\auto@bib@innerbib\@empty
\bibitem [{\citenamefont {Kitaev}(2009)}]{Kitaev2009}%
  \BibitemOpen
  \bibfield  {author} {\bibinfo {author} {\bibfnamefont {A.}~\bibnamefont
  {Kitaev}},\ }\href {\doibase 10.1063/1.3149495} {\bibfield  {journal}
  {\bibinfo  {journal} {AIP Conf. Proc.}\ }\textbf {\bibinfo {volume} {1134}},\
  \bibinfo {pages} {22} (\bibinfo {year} {2009})}\BibitemShut {NoStop}%
\bibitem [{\citenamefont {Altland}\ and\ \citenamefont
  {Zirnbauer}(1997)}]{Altland1997}%
  \BibitemOpen
  \bibfield  {author} {\bibinfo {author} {\bibfnamefont {A.}~\bibnamefont
  {Altland}}\ and\ \bibinfo {author} {\bibfnamefont {M.~R.}\ \bibnamefont
  {Zirnbauer}},\ }\href {\doibase 10.1103/PhysRevB.55.1142} {\bibfield
  {journal} {\bibinfo  {journal} {Phys. Rev. B}\ }\textbf {\bibinfo {volume}
  {55}},\ \bibinfo {pages} {1142} (\bibinfo {year} {1997})}\BibitemShut
  {NoStop}%
\bibitem [{\citenamefont {Thouless}\ \emph {et~al.}(1982)\citenamefont
  {Thouless}, \citenamefont {Kohmoto}, \citenamefont {Nightingale},\ and\
  \citenamefont {denNijs}}]{Thouless1982}%
  \BibitemOpen
  \bibfield  {author} {\bibinfo {author} {\bibfnamefont {D.~J.}\ \bibnamefont
  {Thouless}}, \bibinfo {author} {\bibfnamefont {M.}~\bibnamefont {Kohmoto}},
  \bibinfo {author} {\bibfnamefont {M.~P.}\ \bibnamefont {Nightingale}}, \ and\
  \bibinfo {author} {\bibfnamefont {M.}~\bibnamefont {denNijs}},\ }\href
  {\doibase 10.1103/PhysRevLett.49.405} {\bibfield  {journal} {\bibinfo
  {journal} {Phys. Rev. Lett.}\ }\textbf {\bibinfo {volume} {49}},\ \bibinfo
  {pages} {405} (\bibinfo {year} {1982})}\BibitemShut {NoStop}%
\bibitem [{\citenamefont {Kitaev}(2001)}]{Kitaev2001}%
  \BibitemOpen
  \bibfield  {author} {\bibinfo {author} {\bibfnamefont {A.~Y.}\ \bibnamefont
  {Kitaev}},\ }\href {\doibase 10.1070/1063-7869/44/10S/S29} {\bibfield
  {journal} {\bibinfo  {journal} {Physics-Uspekhi}\ }\textbf {\bibinfo {volume}
  {44}},\ \bibinfo {pages} {131} (\bibinfo {year} {2001})}\BibitemShut
  {NoStop}%
\bibitem [{\citenamefont {Kane}\ and\ \citenamefont {Mele}(2005)}]{Kane2005}%
  \BibitemOpen
  \bibfield  {author} {\bibinfo {author} {\bibfnamefont {C.~L.}\ \bibnamefont
  {Kane}}\ and\ \bibinfo {author} {\bibfnamefont {E.~J.}\ \bibnamefont
  {Mele}},\ }\href {\doibase 10.1103/PhysRevLett.95.146802} {\bibfield
  {journal} {\bibinfo  {journal} {Phys. Rev. Lett.}\ }\textbf {\bibinfo
  {volume} {95}},\ \bibinfo {pages} {146802} (\bibinfo {year}
  {2005})}\BibitemShut {NoStop}%
\bibitem [{\citenamefont {Bernevig}\ \emph {et~al.}(2006)\citenamefont
  {Bernevig}, \citenamefont {Hughes},\ and\ \citenamefont
  {Zhang}}]{Bernevig2006}%
  \BibitemOpen
  \bibfield  {author} {\bibinfo {author} {\bibfnamefont {B.~A.}\ \bibnamefont
  {Bernevig}}, \bibinfo {author} {\bibfnamefont {T.~L.}\ \bibnamefont
  {Hughes}}, \ and\ \bibinfo {author} {\bibfnamefont {S.-C.}\ \bibnamefont
  {Zhang}},\ }\href {\doibase 10.1126/science.1133734} {\bibfield  {journal}
  {\bibinfo  {journal} {Science}\ }\textbf {\bibinfo {volume} {314}},\ \bibinfo
  {pages} {1757} (\bibinfo {year} {2006})}\BibitemShut {NoStop}%
\bibitem [{\citenamefont {Fu}\ \emph {et~al.}(2007)\citenamefont {Fu},
  \citenamefont {Kane},\ and\ \citenamefont {Mele}}]{Fu2007}%
  \BibitemOpen
  \bibfield  {author} {\bibinfo {author} {\bibfnamefont {L.}~\bibnamefont
  {Fu}}, \bibinfo {author} {\bibfnamefont {C.~L.}\ \bibnamefont {Kane}}, \ and\
  \bibinfo {author} {\bibfnamefont {E.~J.}\ \bibnamefont {Mele}},\ }\href
  {\doibase 10.1103/PhysRevLett.98.106803} {\bibfield  {journal} {\bibinfo
  {journal} {Phys. Rev. Lett.}\ }\textbf {\bibinfo {volume} {98}},\ \bibinfo
  {pages} {106803} (\bibinfo {year} {2007})}\BibitemShut {NoStop}%
\bibitem [{\citenamefont {Fu}\ and\ \citenamefont {Kane}(2007)}]{Fu2007b}%
  \BibitemOpen
  \bibfield  {author} {\bibinfo {author} {\bibfnamefont {L.}~\bibnamefont
  {Fu}}\ and\ \bibinfo {author} {\bibfnamefont {C.~L.}\ \bibnamefont {Kane}},\
  }\href {\doibase 10.1103/PhysRevB.76.045302} {\bibfield  {journal} {\bibinfo
  {journal} {Phys. Rev. B}\ }\textbf {\bibinfo {volume} {76}},\ \bibinfo
  {pages} {045302} (\bibinfo {year} {2007})}\BibitemShut {NoStop}%
\bibitem [{\citenamefont {Moore}\ and\ \citenamefont
  {Balents}(2007)}]{Moore2007}%
  \BibitemOpen
  \bibfield  {author} {\bibinfo {author} {\bibfnamefont {J.~E.}\ \bibnamefont
  {Moore}}\ and\ \bibinfo {author} {\bibfnamefont {L.}~\bibnamefont
  {Balents}},\ }\href {\doibase 10.1103/PhysRevB.75.121306} {\bibfield
  {journal} {\bibinfo  {journal} {Phys. Rev. B}\ }\textbf {\bibinfo {volume}
  {75}},\ \bibinfo {pages} {121306(R)} (\bibinfo {year} {2007})}\BibitemShut
  {NoStop}%
\bibitem [{\citenamefont {Roy}(2009)}]{Roy2009}%
  \BibitemOpen
  \bibfield  {author} {\bibinfo {author} {\bibfnamefont {R.}~\bibnamefont
  {Roy}},\ }\href {\doibase 10.1103/PhysRevB.79.195322} {\bibfield  {journal}
  {\bibinfo  {journal} {Phys. Rev. B}\ }\textbf {\bibinfo {volume} {79}},\
  \bibinfo {pages} {195322} (\bibinfo {year} {2009})}\BibitemShut {NoStop}%
\bibitem [{\citenamefont {Fu}(2011)}]{Fu2011}%
  \BibitemOpen
  \bibfield  {author} {\bibinfo {author} {\bibfnamefont {L.}~\bibnamefont
  {Fu}},\ }\href {\doibase 10.1103/PhysRevLett.106.106802} {\bibfield
  {journal} {\bibinfo  {journal} {Phys. Rev. Lett.}\ }\textbf {\bibinfo
  {volume} {106}},\ \bibinfo {pages} {106802} (\bibinfo {year}
  {2011})}\BibitemShut {NoStop}%
\bibitem [{\citenamefont {Benalcazar}\ \emph {et~al.}(2017)\citenamefont
  {Benalcazar}, \citenamefont {Bernevig},\ and\ \citenamefont
  {Hughes}}]{Benalcazar2017}%
  \BibitemOpen
  \bibfield  {author} {\bibinfo {author} {\bibfnamefont {W.~A.}\ \bibnamefont
  {Benalcazar}}, \bibinfo {author} {\bibfnamefont {B.~A.}\ \bibnamefont
  {Bernevig}}, \ and\ \bibinfo {author} {\bibfnamefont {T.~L.}\ \bibnamefont
  {Hughes}},\ }\href {\doibase 10.1103/PhysRevB.96.245115} {\bibfield
  {journal} {\bibinfo  {journal} {Phys. Rev. B}\ }\textbf {\bibinfo {volume}
  {96}},\ \bibinfo {pages} {245115} (\bibinfo {year} {2017})}\BibitemShut
  {NoStop}%
\bibitem [{\citenamefont {Song}\ \emph {et~al.}(2017)\citenamefont {Song},
  \citenamefont {Fang},\ and\ \citenamefont {Fang}}]{Song2017}%
  \BibitemOpen
  \bibfield  {author} {\bibinfo {author} {\bibfnamefont {Z.}~\bibnamefont
  {Song}}, \bibinfo {author} {\bibfnamefont {Z.}~\bibnamefont {Fang}}, \ and\
  \bibinfo {author} {\bibfnamefont {C.}~\bibnamefont {Fang}},\ }\href {\doibase
  10.1103/PhysRevLett.119.246402} {\bibfield  {journal} {\bibinfo  {journal}
  {Phys. Rev. Lett.}\ }\textbf {\bibinfo {volume} {119}},\ \bibinfo {pages}
  {246402} (\bibinfo {year} {2017})}\BibitemShut {NoStop}%
\bibitem [{\citenamefont {Langbehn}\ \emph {et~al.}(2017)\citenamefont
  {Langbehn}, \citenamefont {Peng}, \citenamefont {Trifunovic}, \citenamefont
  {von Oppen},\ and\ \citenamefont {Brouwer}}]{Langbehn2017}%
  \BibitemOpen
  \bibfield  {author} {\bibinfo {author} {\bibfnamefont {J.}~\bibnamefont
  {Langbehn}}, \bibinfo {author} {\bibfnamefont {Y.}~\bibnamefont {Peng}},
  \bibinfo {author} {\bibfnamefont {L.}~\bibnamefont {Trifunovic}}, \bibinfo
  {author} {\bibfnamefont {F.}~\bibnamefont {von Oppen}}, \ and\ \bibinfo
  {author} {\bibfnamefont {P.~W.}\ \bibnamefont {Brouwer}},\ }\href {\doibase
  10.1103/PhysRevLett.119.246401} {\bibfield  {journal} {\bibinfo  {journal}
  {Phys. Rev. Lett.}\ }\textbf {\bibinfo {volume} {119}},\ \bibinfo {pages}
  {246401} (\bibinfo {year} {2017})}\BibitemShut {NoStop}%
\bibitem [{\citenamefont {Khalaf}(2018)}]{Khalaf2018}%
  \BibitemOpen
  \bibfield  {author} {\bibinfo {author} {\bibfnamefont {E.}~\bibnamefont
  {Khalaf}},\ }\href {\doibase 10.1103/PhysRevB.97.205136} {\bibfield
  {journal} {\bibinfo  {journal} {Phys. Rev. B}\ }\textbf {\bibinfo {volume}
  {97}},\ \bibinfo {pages} {205136} (\bibinfo {year} {2018})}\BibitemShut
  {NoStop}%
\bibitem [{\citenamefont {van Miert}\ and\ \citenamefont
  {Ortix}(2018)}]{Ortix2018}%
  \BibitemOpen
  \bibfield  {author} {\bibinfo {author} {\bibfnamefont {G.}~\bibnamefont {van
  Miert}}\ and\ \bibinfo {author} {\bibfnamefont {C.}~\bibnamefont {Ortix}},\
  }\href {\doibase 10.1103/PhysRevB.98.081110} {\bibfield  {journal} {\bibinfo
  {journal} {Phys. Rev. B}\ }\textbf {\bibinfo {volume} {98}},\ \bibinfo
  {pages} {081110(R)} (\bibinfo {year} {2018})}\BibitemShut {NoStop}%
\bibitem [{\citenamefont {Po}\ \emph {et~al.}(2017)\citenamefont {Po},
  \citenamefont {Vishwanath},\ and\ \citenamefont {Watanabe}}]{Po2017}%
  \BibitemOpen
  \bibfield  {author} {\bibinfo {author} {\bibfnamefont {H.~C.}\ \bibnamefont
  {Po}}, \bibinfo {author} {\bibfnamefont {A.}~\bibnamefont {Vishwanath}}, \
  and\ \bibinfo {author} {\bibfnamefont {H.}~\bibnamefont {Watanabe}},\ }\href
  {\doibase 10.1038/s41467-017-00133-2} {\bibfield  {journal} {\bibinfo
  {journal} {Nat. Comm.}\ }\textbf {\bibinfo {volume} {8}},\ \bibinfo {pages}
  {1} (\bibinfo {year} {2017})}\BibitemShut {NoStop}%
\bibitem [{\citenamefont {Ono}\ and\ \citenamefont {Watanabe}(2018)}]{Ono2018}%
  \BibitemOpen
  \bibfield  {author} {\bibinfo {author} {\bibfnamefont {S.}~\bibnamefont
  {Ono}}\ and\ \bibinfo {author} {\bibfnamefont {H.}~\bibnamefont {Watanabe}},\
  }\href {\doibase 10.1103/PhysRevB.98.115150} {\bibfield  {journal} {\bibinfo
  {journal} {Phys. Rev. B}\ }\textbf {\bibinfo {volume} {98}},\ \bibinfo
  {pages} {115150} (\bibinfo {year} {2018})}\BibitemShut {NoStop}%
\bibitem [{\citenamefont {Khalaf}\ \emph {et~al.}(2018)\citenamefont {Khalaf},
  \citenamefont {Po}, \citenamefont {Vishwanath},\ and\ \citenamefont
  {Watanabe}}]{Khalaf2018b}%
  \BibitemOpen
  \bibfield  {author} {\bibinfo {author} {\bibfnamefont {E.}~\bibnamefont
  {Khalaf}}, \bibinfo {author} {\bibfnamefont {H.~C.}\ \bibnamefont {Po}},
  \bibinfo {author} {\bibfnamefont {A.}~\bibnamefont {Vishwanath}}, \ and\
  \bibinfo {author} {\bibfnamefont {H.}~\bibnamefont {Watanabe}},\ }\href
  {\doibase 10.1103/PhysRevX.8.031070} {\bibfield  {journal} {\bibinfo
  {journal} {Phys. Rev. X}\ }\textbf {\bibinfo {volume} {8}},\ \bibinfo {pages}
  {031070} (\bibinfo {year} {2018})}\BibitemShut {NoStop}%
\bibitem [{\citenamefont {Kruthoff}\ \emph {et~al.}(2017)\citenamefont
  {Kruthoff}, \citenamefont {de~Boer}, \citenamefont {van Wezel}, \citenamefont
  {Kane},\ and\ \citenamefont {Slager}}]{Kruthoff2017}%
  \BibitemOpen
  \bibfield  {author} {\bibinfo {author} {\bibfnamefont {J.}~\bibnamefont
  {Kruthoff}}, \bibinfo {author} {\bibfnamefont {J.}~\bibnamefont {de~Boer}},
  \bibinfo {author} {\bibfnamefont {J.}~\bibnamefont {van Wezel}}, \bibinfo
  {author} {\bibfnamefont {C.~L.}\ \bibnamefont {Kane}}, \ and\ \bibinfo
  {author} {\bibfnamefont {R.-J.}\ \bibnamefont {Slager}},\ }\href {\doibase
  10.1103/PhysRevX.7.041069} {\bibfield  {journal} {\bibinfo  {journal} {Phys.
  Rev. X}\ }\textbf {\bibinfo {volume} {7}},\ \bibinfo {pages} {041069}
  (\bibinfo {year} {2017})}\BibitemShut {NoStop}%
\bibitem [{\citenamefont {Yao}\ \emph {et~al.}(2007)\citenamefont {Yao},
  \citenamefont {MacDonald},\ and\ \citenamefont {Niu}}]{Yao2007}%
  \BibitemOpen
  \bibfield  {author} {\bibinfo {author} {\bibfnamefont {W.}~\bibnamefont
  {Yao}}, \bibinfo {author} {\bibfnamefont {A.~H.}\ \bibnamefont {MacDonald}},
  \ and\ \bibinfo {author} {\bibfnamefont {Q.}~\bibnamefont {Niu}},\ }\href
  {\doibase 10.1103/PhysRevLett.99.047401} {\bibfield  {journal} {\bibinfo
  {journal} {Phys. Rev. Lett.}\ }\textbf {\bibinfo {volume} {99}},\ \bibinfo
  {pages} {047401} (\bibinfo {year} {2007})}\BibitemShut {NoStop}%
\bibitem [{\citenamefont {Inoue}\ and\ \citenamefont
  {Tanaka}(2010)}]{Inoue2010}%
  \BibitemOpen
  \bibfield  {author} {\bibinfo {author} {\bibfnamefont {J.-i.}\ \bibnamefont
  {Inoue}}\ and\ \bibinfo {author} {\bibfnamefont {A.}~\bibnamefont {Tanaka}},\
  }\href {\doibase 10.1103/PhysRevLett.105.017401} {\bibfield  {journal}
  {\bibinfo  {journal} {Phys. Rev. Lett.}\ }\textbf {\bibinfo {volume} {105}},\
  \bibinfo {pages} {017401} (\bibinfo {year} {2010})}\BibitemShut {NoStop}%
\bibitem [{\citenamefont {Lindner}\ \emph {et~al.}(2011)\citenamefont
  {Lindner}, \citenamefont {Refael},\ and\ \citenamefont
  {Galitski}}]{Lindner2011}%
  \BibitemOpen
  \bibfield  {author} {\bibinfo {author} {\bibfnamefont {N.~H.}\ \bibnamefont
  {Lindner}}, \bibinfo {author} {\bibfnamefont {G.}~\bibnamefont {Refael}}, \
  and\ \bibinfo {author} {\bibfnamefont {V.}~\bibnamefont {Galitski}},\
  }\href@noop {} {\bibfield  {journal} {\bibinfo  {journal} {Nat. Phys.}\
  }\textbf {\bibinfo {volume} {7}},\ \bibinfo {pages} {490} (\bibinfo {year}
  {2011})}\BibitemShut {NoStop}%
\bibitem [{\citenamefont {Perez-Piskunow}\ \emph {et~al.}(2014)\citenamefont
  {Perez-Piskunow}, \citenamefont {Usaj}, \citenamefont {Balseiro},\ and\
  \citenamefont {FoaTorres}}]{Perez2014}%
  \BibitemOpen
  \bibfield  {author} {\bibinfo {author} {\bibfnamefont {P.~M.}\ \bibnamefont
  {Perez-Piskunow}}, \bibinfo {author} {\bibfnamefont {G.}~\bibnamefont
  {Usaj}}, \bibinfo {author} {\bibfnamefont {C.~A.}\ \bibnamefont {Balseiro}},
  \ and\ \bibinfo {author} {\bibfnamefont {L.~E.~F.}\ \bibnamefont
  {FoaTorres}},\ }\href {\doibase 10.1103/PhysRevB.89.121401} {\bibfield
  {journal} {\bibinfo  {journal} {Phys. Rev. B}\ }\textbf {\bibinfo {volume}
  {89}},\ \bibinfo {pages} {121401(R)} (\bibinfo {year} {2014})}\BibitemShut
  {NoStop}%
\bibitem [{\citenamefont {FoaTorres}\ \emph {et~al.}(2014)\citenamefont
  {FoaTorres}, \citenamefont {Perez-Piskunow}, \citenamefont {Balseiro},\ and\
  \citenamefont {Usaj}}]{Torres2014}%
  \BibitemOpen
  \bibfield  {author} {\bibinfo {author} {\bibfnamefont {L.~E.~F.}\
  \bibnamefont {FoaTorres}}, \bibinfo {author} {\bibfnamefont {P.~M.}\
  \bibnamefont {Perez-Piskunow}}, \bibinfo {author} {\bibfnamefont {C.~A.}\
  \bibnamefont {Balseiro}}, \ and\ \bibinfo {author} {\bibfnamefont
  {G.}~\bibnamefont {Usaj}},\ }\href {\doibase 10.1103/PhysRevLett.113.266801}
  {\bibfield  {journal} {\bibinfo  {journal} {Phys. Rev. Lett.}\ }\textbf
  {\bibinfo {volume} {113}},\ \bibinfo {pages} {266801} (\bibinfo {year}
  {2014})}\BibitemShut {NoStop}%
\bibitem [{\citenamefont {Rechtsman}\ \emph {et~al.}(2013)\citenamefont
  {Rechtsman}, \citenamefont {Zeuner}, \citenamefont {Plotnik}, \citenamefont
  {Lumer}, \citenamefont {Podolsky}, \citenamefont {Dreisow}, \citenamefont
  {Nolte}, \citenamefont {Segev},\ and\ \citenamefont
  {Szameit}}]{Rechtsman2013}%
  \BibitemOpen
  \bibfield  {author} {\bibinfo {author} {\bibfnamefont {M.~C.}\ \bibnamefont
  {Rechtsman}}, \bibinfo {author} {\bibfnamefont {J.~M.}\ \bibnamefont
  {Zeuner}}, \bibinfo {author} {\bibfnamefont {Y.}~\bibnamefont {Plotnik}},
  \bibinfo {author} {\bibfnamefont {Y.}~\bibnamefont {Lumer}}, \bibinfo
  {author} {\bibfnamefont {D.}~\bibnamefont {Podolsky}}, \bibinfo {author}
  {\bibfnamefont {F.}~\bibnamefont {Dreisow}}, \bibinfo {author} {\bibfnamefont
  {S.}~\bibnamefont {Nolte}}, \bibinfo {author} {\bibfnamefont
  {M.}~\bibnamefont {Segev}}, \ and\ \bibinfo {author} {\bibfnamefont
  {A.}~\bibnamefont {Szameit}},\ }\href@noop {} {\bibfield  {journal} {\bibinfo
   {journal} {Nature}\ }\textbf {\bibinfo {volume} {496}},\ \bibinfo {pages}
  {196} (\bibinfo {year} {2013})}\BibitemShut {NoStop}%
\bibitem [{\citenamefont {Maczewsky}\ \emph {et~al.}(2017)\citenamefont
  {Maczewsky}, \citenamefont {Zeuner}, \citenamefont {Nolte},\ and\
  \citenamefont {Szameit}}]{Maczewsky2017}%
  \BibitemOpen
  \bibfield  {author} {\bibinfo {author} {\bibfnamefont {L.~J.}\ \bibnamefont
  {Maczewsky}}, \bibinfo {author} {\bibfnamefont {J.~M.}\ \bibnamefont
  {Zeuner}}, \bibinfo {author} {\bibfnamefont {S.}~\bibnamefont {Nolte}}, \
  and\ \bibinfo {author} {\bibfnamefont {A.}~\bibnamefont {Szameit}},\
  }\href@noop {} {\bibfield  {journal} {\bibinfo  {journal} {Nature
  communications}\ }\textbf {\bibinfo {volume} {8}},\ \bibinfo {pages} {1}
  (\bibinfo {year} {2017})}\BibitemShut {NoStop}%
\bibitem [{\citenamefont {Jotzu}\ \emph {et~al.}(2014)\citenamefont {Jotzu},
  \citenamefont {Messer}, \citenamefont {Desbuquois}, \citenamefont {Lebrat},
  \citenamefont {Uehlinger}, \citenamefont {Greif},\ and\ \citenamefont
  {Esslinger}}]{Jotzu2014}%
  \BibitemOpen
  \bibfield  {author} {\bibinfo {author} {\bibfnamefont {G.}~\bibnamefont
  {Jotzu}}, \bibinfo {author} {\bibfnamefont {M.}~\bibnamefont {Messer}},
  \bibinfo {author} {\bibfnamefont {R.}~\bibnamefont {Desbuquois}}, \bibinfo
  {author} {\bibfnamefont {M.}~\bibnamefont {Lebrat}}, \bibinfo {author}
  {\bibfnamefont {T.}~\bibnamefont {Uehlinger}}, \bibinfo {author}
  {\bibfnamefont {D.}~\bibnamefont {Greif}}, \ and\ \bibinfo {author}
  {\bibfnamefont {T.}~\bibnamefont {Esslinger}},\ }\href@noop {} {\bibfield
  {journal} {\bibinfo  {journal} {Nature}\ }\textbf {\bibinfo {volume} {515}},\
  \bibinfo {pages} {237} (\bibinfo {year} {2014})}\BibitemShut {NoStop}%
\bibitem [{\citenamefont {Jim\'enez-Garc\'{\i}a}\ \emph
  {et~al.}(2015)\citenamefont {Jim\'enez-Garc\'{\i}a}, \citenamefont {LeBlanc},
  \citenamefont {Williams}, \citenamefont {Beeler}, \citenamefont {Qu},
  \citenamefont {Gong}, \citenamefont {Zhang},\ and\ \citenamefont
  {Spielman}}]{Jim2015}%
  \BibitemOpen
  \bibfield  {author} {\bibinfo {author} {\bibfnamefont {K.}~\bibnamefont
  {Jim\'enez-Garc\'{\i}a}}, \bibinfo {author} {\bibfnamefont {L.~J.}\
  \bibnamefont {LeBlanc}}, \bibinfo {author} {\bibfnamefont {R.~A.}\
  \bibnamefont {Williams}}, \bibinfo {author} {\bibfnamefont {M.~C.}\
  \bibnamefont {Beeler}}, \bibinfo {author} {\bibfnamefont {C.}~\bibnamefont
  {Qu}}, \bibinfo {author} {\bibfnamefont {M.}~\bibnamefont {Gong}}, \bibinfo
  {author} {\bibfnamefont {C.}~\bibnamefont {Zhang}}, \ and\ \bibinfo {author}
  {\bibfnamefont {I.~B.}\ \bibnamefont {Spielman}},\ }\href {\doibase
  10.1103/PhysRevLett.114.125301} {\bibfield  {journal} {\bibinfo  {journal}
  {Phys. Rev. Lett.}\ }\textbf {\bibinfo {volume} {114}},\ \bibinfo {pages}
  {125301} (\bibinfo {year} {2015})}\BibitemShut {NoStop}%
\bibitem [{\citenamefont {Yang}\ \emph {et~al.}(2018)\citenamefont {Yang},
  \citenamefont {Huang},\ and\ \citenamefont {Wang}}]{Yang2018}%
  \BibitemOpen
  \bibfield  {author} {\bibinfo {author} {\bibfnamefont {X.}~\bibnamefont
  {Yang}}, \bibinfo {author} {\bibfnamefont {B.}~\bibnamefont {Huang}}, \ and\
  \bibinfo {author} {\bibfnamefont {Z.}~\bibnamefont {Wang}},\ }\href@noop {}
  {\bibfield  {journal} {\bibinfo  {journal} {Scientific reports}\ }\textbf
  {\bibinfo {volume} {8}},\ \bibinfo {pages} {1} (\bibinfo {year}
  {2018})}\BibitemShut {NoStop}%
\bibitem [{\citenamefont {Kitagawa}\ \emph {et~al.}(2010)\citenamefont
  {Kitagawa}, \citenamefont {Berg}, \citenamefont {Rudner},\ and\ \citenamefont
  {Demler}}]{Kitagawa2010}%
  \BibitemOpen
  \bibfield  {author} {\bibinfo {author} {\bibfnamefont {T.}~\bibnamefont
  {Kitagawa}}, \bibinfo {author} {\bibfnamefont {E.}~\bibnamefont {Berg}},
  \bibinfo {author} {\bibfnamefont {M.}~\bibnamefont {Rudner}}, \ and\ \bibinfo
  {author} {\bibfnamefont {E.}~\bibnamefont {Demler}},\ }\href {\doibase
  10.1103/PhysRevB.82.235114} {\bibfield  {journal} {\bibinfo  {journal} {Phys.
  Rev. B}\ }\textbf {\bibinfo {volume} {82}},\ \bibinfo {pages} {235114}
  (\bibinfo {year} {2010})}\BibitemShut {NoStop}%
\bibitem [{\citenamefont {Jiang}\ \emph {et~al.}(2011)\citenamefont {Jiang},
  \citenamefont {Kitagawa}, \citenamefont {Alicea}, \citenamefont {Akhmerov},
  \citenamefont {Pekker}, \citenamefont {Refael}, \citenamefont {Cirac},
  \citenamefont {Demler}, \citenamefont {Lukin},\ and\ \citenamefont
  {Zoller}}]{Jiang2011}%
  \BibitemOpen
  \bibfield  {author} {\bibinfo {author} {\bibfnamefont {L.}~\bibnamefont
  {Jiang}}, \bibinfo {author} {\bibfnamefont {T.}~\bibnamefont {Kitagawa}},
  \bibinfo {author} {\bibfnamefont {J.}~\bibnamefont {Alicea}}, \bibinfo
  {author} {\bibfnamefont {A.~R.}\ \bibnamefont {Akhmerov}}, \bibinfo {author}
  {\bibfnamefont {D.}~\bibnamefont {Pekker}}, \bibinfo {author} {\bibfnamefont
  {G.}~\bibnamefont {Refael}}, \bibinfo {author} {\bibfnamefont {J.~I.}\
  \bibnamefont {Cirac}}, \bibinfo {author} {\bibfnamefont {E.}~\bibnamefont
  {Demler}}, \bibinfo {author} {\bibfnamefont {M.~D.}\ \bibnamefont {Lukin}}, \
  and\ \bibinfo {author} {\bibfnamefont {P.}~\bibnamefont {Zoller}},\ }\href
  {\doibase 10.1103/PhysRevLett.106.220402} {\bibfield  {journal} {\bibinfo
  {journal} {Phys. Rev. Lett.}\ }\textbf {\bibinfo {volume} {106}},\ \bibinfo
  {pages} {220402} (\bibinfo {year} {2011})}\BibitemShut {NoStop}%
\bibitem [{\citenamefont {Rudner}\ \emph {et~al.}(2013)\citenamefont {Rudner},
  \citenamefont {Lindner}, \citenamefont {Berg},\ and\ \citenamefont
  {Levin}}]{Rudner2013}%
  \BibitemOpen
  \bibfield  {author} {\bibinfo {author} {\bibfnamefont {M.~S.}\ \bibnamefont
  {Rudner}}, \bibinfo {author} {\bibfnamefont {N.~H.}\ \bibnamefont {Lindner}},
  \bibinfo {author} {\bibfnamefont {E.}~\bibnamefont {Berg}}, \ and\ \bibinfo
  {author} {\bibfnamefont {M.}~\bibnamefont {Levin}},\ }\href {\doibase
  10.1103/PhysRevX.3.031005} {\bibfield  {journal} {\bibinfo  {journal} {Phys.
  Rev. X}\ }\textbf {\bibinfo {volume} {3}},\ \bibinfo {pages} {031005}
  (\bibinfo {year} {2013})}\BibitemShut {NoStop}%
\bibitem [{\citenamefont {Nathan}\ and\ \citenamefont
  {Rudner}(2015)}]{Nathan2015}%
  \BibitemOpen
  \bibfield  {author} {\bibinfo {author} {\bibfnamefont {F.}~\bibnamefont
  {Nathan}}\ and\ \bibinfo {author} {\bibfnamefont {M.~S.}\ \bibnamefont
  {Rudner}},\ }\href {\doibase 10.1088/1367-2630/17/12/125014} {\bibfield
  {journal} {\bibinfo  {journal} {New J. Phys.}\ }\textbf {\bibinfo {volume}
  {17}},\ \bibinfo {pages} {125014} (\bibinfo {year} {2015})}\BibitemShut
  {NoStop}%
\bibitem [{\citenamefont {Zhu}\ \emph {et~al.}(2020)\citenamefont {Zhu},
  \citenamefont {Qin}, \citenamefont {Yang}, \citenamefont {Xianlong},\ and\
  \citenamefont {Liang}}]{Zhu2020}%
  \BibitemOpen
  \bibfield  {author} {\bibinfo {author} {\bibfnamefont {Y.}~\bibnamefont
  {Zhu}}, \bibinfo {author} {\bibfnamefont {T.}~\bibnamefont {Qin}}, \bibinfo
  {author} {\bibfnamefont {X.}~\bibnamefont {Yang}}, \bibinfo {author}
  {\bibfnamefont {G.}~\bibnamefont {Xianlong}}, \ and\ \bibinfo {author}
  {\bibfnamefont {Z.}~\bibnamefont {Liang}},\ }\href {\doibase
  10.1103/PhysRevResearch.2.033045} {\bibfield  {journal} {\bibinfo  {journal}
  {Phys. Rev. Research}\ }\textbf {\bibinfo {volume} {2}},\ \bibinfo {pages}
  {033045} (\bibinfo {year} {2020})}\BibitemShut {NoStop}%
\bibitem [{\citenamefont {Yu}\ \emph {et~al.}(2021)\citenamefont {Yu},
  \citenamefont {Zhang},\ and\ \citenamefont {Song}}]{Yu2021}%
  \BibitemOpen
  \bibfield  {author} {\bibinfo {author} {\bibfnamefont {J.}~\bibnamefont
  {Yu}}, \bibinfo {author} {\bibfnamefont {R.-X.}\ \bibnamefont {Zhang}}, \
  and\ \bibinfo {author} {\bibfnamefont {Z.-D.}\ \bibnamefont {Song}},\ }\href
  {\doibase 10.1038/s41467-021-26092-3} {\bibfield  {journal} {\bibinfo
  {journal} {Nat. Commun.}\ }\textbf {\bibinfo {volume} {12}},\ \bibinfo
  {pages} {5985} (\bibinfo {year} {2021})}\BibitemShut {NoStop}%
\bibitem [{\citenamefont {Roy}\ and\ \citenamefont {Harper}(2017)}]{Roy2017}%
  \BibitemOpen
  \bibfield  {author} {\bibinfo {author} {\bibfnamefont {R.}~\bibnamefont
  {Roy}}\ and\ \bibinfo {author} {\bibfnamefont {F.}~\bibnamefont {Harper}},\
  }\href {\doibase 10.1103/PhysRevB.96.155118} {\bibfield  {journal} {\bibinfo
  {journal} {Phys. Rev. B}\ }\textbf {\bibinfo {volume} {96}},\ \bibinfo
  {pages} {155118} (\bibinfo {year} {2017})}\BibitemShut {NoStop}%
\bibitem [{\citenamefont {Peng}(2020)}]{Peng2020}%
  \BibitemOpen
  \bibfield  {author} {\bibinfo {author} {\bibfnamefont {Y.}~\bibnamefont
  {Peng}},\ }\href {\doibase 10.1103/PhysRevResearch.2.013124} {\bibfield
  {journal} {\bibinfo  {journal} {Phys. Rev. Res.}\ }\textbf {\bibinfo {volume}
  {2}},\ \bibinfo {pages} {013124} (\bibinfo {year} {2020})}\BibitemShut
  {NoStop}%
\bibitem [{\citenamefont {Fruchart}(2016)}]{Fruchart2016}%
  \BibitemOpen
  \bibfield  {author} {\bibinfo {author} {\bibfnamefont {M.}~\bibnamefont
  {Fruchart}},\ }\href {\doibase 10.1103/PhysRevB.93.115429} {\bibfield
  {journal} {\bibinfo  {journal} {Phys. Rev. B}\ }\textbf {\bibinfo {volume}
  {93}},\ \bibinfo {pages} {115429} (\bibinfo {year} {2016})}\BibitemShut
  {NoStop}%
\bibitem [{\citenamefont {Asb\'oth}\ \emph {et~al.}(2014)\citenamefont
  {Asb\'oth}, \citenamefont {Tarasinski},\ and\ \citenamefont
  {Delplace}}]{Asb2014}%
  \BibitemOpen
  \bibfield  {author} {\bibinfo {author} {\bibfnamefont {J.~K.}\ \bibnamefont
  {Asb\'oth}}, \bibinfo {author} {\bibfnamefont {B.}~\bibnamefont
  {Tarasinski}}, \ and\ \bibinfo {author} {\bibfnamefont {P.}~\bibnamefont
  {Delplace}},\ }\href {\doibase 10.1103/PhysRevB.90.125143} {\bibfield
  {journal} {\bibinfo  {journal} {Phys. Rev. B}\ }\textbf {\bibinfo {volume}
  {90}},\ \bibinfo {pages} {125143} (\bibinfo {year} {2014})}\BibitemShut
  {NoStop}%
\bibitem [{\citenamefont {Zhang}\ and\ \citenamefont {Yang}(2020)}]{Zhang2020}%
  \BibitemOpen
  \bibfield  {author} {\bibinfo {author} {\bibfnamefont {R.-X.}\ \bibnamefont
  {Zhang}}\ and\ \bibinfo {author} {\bibfnamefont {Z.-C.}\ \bibnamefont
  {Yang}},\ }\href {http://arxiv.org/abs/2010.07945} {\enquote {\bibinfo
  {title} {{Theory of Anomalous Floquet Higher-Order Topology: Classification,
  Characterization, and Bulk-Boundary Correspondence}},}\ } (\bibinfo {year}
  {2020}),\ \Eprint {http://arxiv.org/abs/2010.07945} {2010.07945} \BibitemShut
  {NoStop}%
\bibitem [{\citenamefont {Zhu}\ \emph {et~al.}(2021)\citenamefont {Zhu},
  \citenamefont {Chong},\ and\ \citenamefont {Gong}}]{Zhu2021}%
  \BibitemOpen
  \bibfield  {author} {\bibinfo {author} {\bibfnamefont {W.}~\bibnamefont
  {Zhu}}, \bibinfo {author} {\bibfnamefont {Y.~D.}\ \bibnamefont {Chong}}, \
  and\ \bibinfo {author} {\bibfnamefont {J.}~\bibnamefont {Gong}},\ }\href
  {\doibase 10.1103/PhysRevB.104.L020302} {\bibfield  {journal} {\bibinfo
  {journal} {Phys. Rev. B}\ }\textbf {\bibinfo {volume} {104}},\ \bibinfo
  {pages} {L020302} (\bibinfo {year} {2021})}\BibitemShut {NoStop}%
\bibitem [{\citenamefont {Carpentier}\ \emph {et~al.}(2015)\citenamefont
  {Carpentier}, \citenamefont {Delplace}, \citenamefont {Fruchart},\ and\
  \citenamefont {Gaw{\c{e}}dzki}}]{Carpentier2015}%
  \BibitemOpen
  \bibfield  {author} {\bibinfo {author} {\bibfnamefont {D.}~\bibnamefont
  {Carpentier}}, \bibinfo {author} {\bibfnamefont {P.}~\bibnamefont
  {Delplace}}, \bibinfo {author} {\bibfnamefont {M.}~\bibnamefont {Fruchart}},
  \ and\ \bibinfo {author} {\bibfnamefont {K.}~\bibnamefont {Gaw{\c{e}}dzki}},\
  }\href {\doibase 10.1103/PhysRevLett.114.106806} {\bibfield  {journal}
  {\bibinfo  {journal} {Phys. Rev. Lett.}\ }\textbf {\bibinfo {volume} {114}},\
  \bibinfo {pages} {106806} (\bibinfo {year} {2015})}\BibitemShut {NoStop}%
\bibitem [{\citenamefont {Vu}\ \emph {et~al.}(2021)\citenamefont {Vu},
  \citenamefont {Zhang}, \citenamefont {Yang},\ and\ \citenamefont
  {Das~Sarma}}]{Vu2021}%
  \BibitemOpen
  \bibfield  {author} {\bibinfo {author} {\bibfnamefont {D.~D.}\ \bibnamefont
  {Vu}}, \bibinfo {author} {\bibfnamefont {R.-X.}\ \bibnamefont {Zhang}},
  \bibinfo {author} {\bibfnamefont {Z.-C.}\ \bibnamefont {Yang}}, \ and\
  \bibinfo {author} {\bibfnamefont {S.}~\bibnamefont {Das~Sarma}},\ }\href
  {\doibase 10.1103/PhysRevB.104.L140502} {\bibfield  {journal} {\bibinfo
  {journal} {Phys. Rev. B}\ }\textbf {\bibinfo {volume} {104}},\ \bibinfo
  {pages} {L140502} (\bibinfo {year} {2021})}\BibitemShut {NoStop}%
\bibitem [{\citenamefont {Shiozaki}\ \emph {et~al.}(2018)\citenamefont
  {Shiozaki}, \citenamefont {Sato},\ and\ \citenamefont {Gomi}}]{Shiozaki2018}%
  \BibitemOpen
  \bibfield  {author} {\bibinfo {author} {\bibfnamefont {K.}~\bibnamefont
  {Shiozaki}}, \bibinfo {author} {\bibfnamefont {M.}~\bibnamefont {Sato}}, \
  and\ \bibinfo {author} {\bibfnamefont {K.}~\bibnamefont {Gomi}},\ }\href
  {http://arxiv.org/abs/1802.06694} {\enquote {\bibinfo {title}
  {{Atiyah-Hirzebruch Spectral Sequence in Band Topology: General Formalism and
  Topological Invariants for 230 Space Groups}},}\ } (\bibinfo {year} {2018}),\
  \Eprint {http://arxiv.org/abs/1802.06694} {arXiv:1802.06694} \BibitemShut
  {NoStop}%
\bibitem [{\citenamefont {Stehouwer}\ \emph {et~al.}(2018)\citenamefont
  {Stehouwer}, \citenamefont {de~Boer}, \citenamefont {Kruthoff},\ and\
  \citenamefont {Posthuma}}]{Stehouwer2018}%
  \BibitemOpen
  \bibfield  {author} {\bibinfo {author} {\bibfnamefont {L.}~\bibnamefont
  {Stehouwer}}, \bibinfo {author} {\bibfnamefont {J.}~\bibnamefont {de~Boer}},
  \bibinfo {author} {\bibfnamefont {J.}~\bibnamefont {Kruthoff}}, \ and\
  \bibinfo {author} {\bibfnamefont {H.}~\bibnamefont {Posthuma}},\ }\href@noop
  {} {\enquote {\bibinfo {title} {Classification of crystalline topological
  insulators through k-theory},}\ } (\bibinfo {year} {2018}),\ \Eprint
  {http://arxiv.org/abs/1811.02592} {arXiv:1811.02592} \BibitemShut {NoStop}%
\bibitem [{\citenamefont {Okuma}\ \emph {et~al.}(2019)\citenamefont {Okuma},
  \citenamefont {Sato},\ and\ \citenamefont {Shiozaki}}]{Okuma2019}%
  \BibitemOpen
  \bibfield  {author} {\bibinfo {author} {\bibfnamefont {N.}~\bibnamefont
  {Okuma}}, \bibinfo {author} {\bibfnamefont {M.}~\bibnamefont {Sato}}, \ and\
  \bibinfo {author} {\bibfnamefont {K.}~\bibnamefont {Shiozaki}},\ }\href
  {\doibase 10.1103/PhysRevB.99.085127} {\bibfield  {journal} {\bibinfo
  {journal} {Phys. Rev. B}\ }\textbf {\bibinfo {volume} {99}},\ \bibinfo
  {pages} {085127} (\bibinfo {year} {2019})}\BibitemShut {NoStop}%
\bibitem [{\citenamefont {Huang}\ and\ \citenamefont {Hsu}(2021)}]{Huang2021}%
  \BibitemOpen
  \bibfield  {author} {\bibinfo {author} {\bibfnamefont {S.-J.}\ \bibnamefont
  {Huang}}\ and\ \bibinfo {author} {\bibfnamefont {Y.-T.}\ \bibnamefont
  {Hsu}},\ }\href@noop {} {\bibfield  {journal} {\bibinfo  {journal} {Phys.
  Rev. Res.}\ }\textbf {\bibinfo {volume} {3}},\ \bibinfo {pages} {013243}
  (\bibinfo {year} {2021})}\BibitemShut {NoStop}%
\bibitem [{\citenamefont {Chen}\ \emph {et~al.}(2021)\citenamefont {Chen},
  \citenamefont {Huang}, \citenamefont {Hsu},\ and\ \citenamefont
  {Wei}}]{Chen2021}%
  \BibitemOpen
  \bibfield  {author} {\bibinfo {author} {\bibfnamefont {Y.}~\bibnamefont
  {Chen}}, \bibinfo {author} {\bibfnamefont {S.-J.}\ \bibnamefont {Huang}},
  \bibinfo {author} {\bibfnamefont {Y.-T.}\ \bibnamefont {Hsu}}, \ and\
  \bibinfo {author} {\bibfnamefont {T.-C.}\ \bibnamefont {Wei}},\ }\href
  {http://arxiv.org/abs/2109.06959} {\enquote {\bibinfo {title}
  {{Boundary-diagnosing topological invariants beyond symmetry indicators: A
  case study of two-fold rotational symmetric superconductors}},}\ } (\bibinfo
  {year} {2021}),\ \Eprint {http://arxiv.org/abs/2109.06959} {arXiv:2109.06959}
  \BibitemShut {NoStop}%
\bibitem [{\citenamefont {Qi}\ \emph {et~al.}(2008)\citenamefont {Qi},
  \citenamefont {Hughes},\ and\ \citenamefont {Zhang}}]{Qi2008}%
  \BibitemOpen
  \bibfield  {author} {\bibinfo {author} {\bibfnamefont {X.~L.}\ \bibnamefont
  {Qi}}, \bibinfo {author} {\bibfnamefont {T.~L.}\ \bibnamefont {Hughes}}, \
  and\ \bibinfo {author} {\bibfnamefont {S.~C.}\ \bibnamefont {Zhang}},\ }\href
  {\doibase 10.1103/PhysRevB.78.195424} {\bibfield  {journal} {\bibinfo
  {journal} {Phys. Rev. B}\ }\textbf {\bibinfo {volume} {78}},\ \bibinfo
  {pages} {195424} (\bibinfo {year} {2008})}\BibitemShut {NoStop}%
\bibitem [{\citenamefont {Ryu}\ \emph {et~al.}(2010)\citenamefont {Ryu},
  \citenamefont {Schnyder}, \citenamefont {Furusaki},\ and\ \citenamefont
  {Ludwig}}]{Ryu2010}%
  \BibitemOpen
  \bibfield  {author} {\bibinfo {author} {\bibfnamefont {S.}~\bibnamefont
  {Ryu}}, \bibinfo {author} {\bibfnamefont {A.~P.}\ \bibnamefont {Schnyder}},
  \bibinfo {author} {\bibfnamefont {A.}~\bibnamefont {Furusaki}}, \ and\
  \bibinfo {author} {\bibfnamefont {A.~W.~W.}\ \bibnamefont {Ludwig}},\ }\href
  {\doibase 10.1088/1367-2630/12/6/065010} {\bibfield  {journal} {\bibinfo
  {journal} {New J. Phys.}\ }\textbf {\bibinfo {volume} {12}},\ \bibinfo
  {pages} {065010} (\bibinfo {year} {2010})}\BibitemShut {NoStop}%
\bibitem [{\citenamefont {Wang}\ \emph {et~al.}(2010)\citenamefont {Wang},
  \citenamefont {Qi},\ and\ \citenamefont {Zhang}}]{Wang2010}%
  \BibitemOpen
  \bibfield  {author} {\bibinfo {author} {\bibfnamefont {Z.}~\bibnamefont
  {Wang}}, \bibinfo {author} {\bibfnamefont {X.-L.}\ \bibnamefont {Qi}}, \ and\
  \bibinfo {author} {\bibfnamefont {S.-C.}\ \bibnamefont {Zhang}},\ }\href@noop
  {} {\bibfield  {journal} {\bibinfo  {journal} {New Journal of Physics}\
  }\textbf {\bibinfo {volume} {12}},\ \bibinfo {pages} {065007} (\bibinfo
  {year} {2010})}\BibitemShut {NoStop}%
\bibitem [{\citenamefont {Turner}\ \emph {et~al.}(2012)\citenamefont {Turner},
  \citenamefont {Zhang}, \citenamefont {Mong},\ and\ \citenamefont
  {Vishwanath}}]{Turner2012}%
  \BibitemOpen
  \bibfield  {author} {\bibinfo {author} {\bibfnamefont {A.~M.}\ \bibnamefont
  {Turner}}, \bibinfo {author} {\bibfnamefont {Y.}~\bibnamefont {Zhang}},
  \bibinfo {author} {\bibfnamefont {R.~S.~K.}\ \bibnamefont {Mong}}, \ and\
  \bibinfo {author} {\bibfnamefont {A.}~\bibnamefont {Vishwanath}},\ }\href
  {\doibase 10.1103/PhysRevB.85.165120} {\bibfield  {journal} {\bibinfo
  {journal} {Phys. Rev. B}\ }\textbf {\bibinfo {volume} {85}},\ \bibinfo
  {pages} {165120} (\bibinfo {year} {2012})}\BibitemShut {NoStop}%
\bibitem [{\citenamefont {Fang}\ \emph {et~al.}(2012)\citenamefont {Fang},
  \citenamefont {Gilbert},\ and\ \citenamefont {Bernevig}}]{Fang2012}%
  \BibitemOpen
  \bibfield  {author} {\bibinfo {author} {\bibfnamefont {C.}~\bibnamefont
  {Fang}}, \bibinfo {author} {\bibfnamefont {M.~J.}\ \bibnamefont {Gilbert}}, \
  and\ \bibinfo {author} {\bibfnamefont {B.~A.}\ \bibnamefont {Bernevig}},\
  }\href {\doibase 10.1103/PhysRevB.86.115112} {\bibfield  {journal} {\bibinfo
  {journal} {Phys. Rev. B}\ }\textbf {\bibinfo {volume} {86}},\ \bibinfo
  {pages} {115112} (\bibinfo {year} {2012})}\BibitemShut {NoStop}%
\bibitem [{\citenamefont {Wieder}\ and\ \citenamefont
  {Bernevig}(2018)}]{Wieder2018}%
  \BibitemOpen
  \bibfield  {author} {\bibinfo {author} {\bibfnamefont {B.~J.}\ \bibnamefont
  {Wieder}}\ and\ \bibinfo {author} {\bibfnamefont {B.~A.}\ \bibnamefont
  {Bernevig}},\ }\href@noop {} {\enquote {\bibinfo {title} {The axion insulator
  as a pump of fragile topology},}\ } (\bibinfo {year} {2018}),\ \Eprint
  {http://arxiv.org/abs/1810.02373} {arXiv:1810.02373 [cond-mat.mes-hall]}
  \BibitemShut {NoStop}%
\bibitem [{\citenamefont {Yu}\ \emph {et~al.}(2020)\citenamefont {Yu},
  \citenamefont {Song},\ and\ \citenamefont {Liu}}]{Yu2020}%
  \BibitemOpen
  \bibfield  {author} {\bibinfo {author} {\bibfnamefont {J.}~\bibnamefont
  {Yu}}, \bibinfo {author} {\bibfnamefont {Z.-D.}\ \bibnamefont {Song}}, \ and\
  \bibinfo {author} {\bibfnamefont {C.-X.}\ \bibnamefont {Liu}},\ }\href
  {\doibase 10.1103/PhysRevLett.125.036401} {\bibfield  {journal} {\bibinfo
  {journal} {Phys. Rev. Lett.}\ }\textbf {\bibinfo {volume} {125}},\ \bibinfo
  {pages} {036401} (\bibinfo {year} {2020})}\BibitemShut {NoStop}%
\bibitem [{\citenamefont {Zhang}\ and\ \citenamefont {{Das
  Sarma}}(2021)}]{Zhang2021}%
  \BibitemOpen
  \bibfield  {author} {\bibinfo {author} {\bibfnamefont {R.~X.}\ \bibnamefont
  {Zhang}}\ and\ \bibinfo {author} {\bibfnamefont {S.}~\bibnamefont {{Das
  Sarma}}},\ }\href {\doibase 10.1103/PhysRevLett.127.067001} {\bibfield
  {journal} {\bibinfo  {journal} {Phys. Rev. Lett.}\ }\textbf {\bibinfo
  {volume} {127}},\ \bibinfo {pages} {067001} (\bibinfo {year}
  {2021})}\BibitemShut {NoStop}%
\end{thebibliography}%
	
	\appendix
	\section{Building block of the winding number}
	In this section, we prove that in a $d-$dimension BZ with odd $d=2n+1$, a structure described by
	\begin{equation}
	U(\kk) = \begin{cases}
	-\exp\left[i(k_i O_{i,j} \gamma_j)\right] & \left|\hspace{0.03in}\sum_j \left(k_iO_{i,j}\right)^2 \le \pi^2 \right. \\
	\hspace{0.5in}\mathbbm{1} &\text{  otherwise }
	\end{cases}
	\end{equation} 
	has the winding number $\pm1$ and thus can be regarded as the building block for the non-trivial winding number. Due to the anti-commutation between Clifford generators $\gamma$, the exponenet can be expressed as
	\begin{equation}
	\begin{split}
	\exp\left[i\left(k_i O_{i,j} \gamma_j\right)\right] = \mathbbm{1}\cos\theta + \frac{i\sin\theta}{\theta}\theta_j\gamma_j,
	\end{split}
	\end{equation} 
	where $\theta_j = k_iO_{i,j}$ and $\theta=\sqrt{\sum \theta_j^2}$ with $0\le\theta\le\pi$. We can further define a vector $\mathbf{A}=(A^0,A^1,\dots,A^d)^T$ with $A^0=\cos\theta$ and $A^j = \sin\theta/\theta$. We note that $A^\alpha A^\alpha = 1$, defining a $d-$dimension hypersphere. The new set of matrices are also defined by $\tilde{\gamma}_0=\mathbbm{1}$ and $\tilde{\gamma}_j=i\gamma_j$ so that $\tilde{\gamma}_i\tilde{\gamma}_j^\dagger=-\tilde{\gamma}_j\tilde{\gamma}_i^\dagger$. We first need manipulate the intergrand
		\begin{equation}
		\begin{split}
		&\text{Tr}\left[(U^\dagger\partial_{k_{\alpha_1}}U)(U^\dagger\partial_{k_{\alpha_2}}U)\dots (U^\dagger\partial_{k_{\alpha_d}}U)\right]\\  &= \text{Tr}\left[U^\dagger\partial_{k_{\alpha_1}}U\partial_{k_{\alpha_2}}U^\dagger\dots\partial_{k_{\alpha_d}}U\right]\\
		&= \text{Tr}\left[\tilde{\gamma}_{\beta_0}^\dagger \tilde{\gamma}_{\beta_1}\tilde{\gamma}_{\beta_2}^\dagger\dots\tilde{\gamma}_{\beta_d}\right]A^{\beta_0}\partial_{k_{\alpha_1}}A^{\beta_1}\partial_{k_{\alpha_1}}A^{\beta_1}\dots\partial_{k_{\alpha_d}}A^{\beta_d}\\
		&=2^n i^{3n+1} \epsilon_{\beta_0\beta_1\beta_2\dots\beta_d}\left(A^{\beta_0}\partial_{k_{\alpha_1}}A^{\beta_1}\partial_{k_{\alpha_1}}A^{\beta_1}\dots\partial_{k_{\alpha_d}}A^{\beta_d}\right) \\
		&= i(-2i)^n  \text{Det}[\mathbf{A},\partial_{k_{\alpha_1}}\mathbf{A},\partial_{k_{\alpha_2}}\mathbf{A},\dots,\partial_{k_{\alpha_d}}\mathbf{A}]\\
		&= i(-2i)^n  \text{Det}[\mathbf{A},\partial_{\theta_{\alpha_1}}\mathbf{A},\partial_{\theta_{\alpha_2}}\mathbf{A},\dots,\partial_{\theta_{\alpha_d}}\mathbf{A}]  \text{Det}[O]
		\end{split}
		\end{equation}
		The value of the $\text{Tr}$ is obtained inductively from the ``anti-commutation" and fact that $2A^\alpha\partial_\beta A^\alpha = \partial_\beta(A^\alpha A^\alpha)=0$. On the hand, $\int d^dk = \int d\theta^d \left| \text{Det}[O] \right|^{-1}$, allowing us to rewrite the winding number integral as
		\begin{equation}
		\begin{split}
		&\int_{BZ}\epsilon^{\alpha_1,\alpha_2,\dots,\alpha_d} \text{Tr}\left[(U^\dagger\partial_{k_{\alpha_1}}U)(U^\dagger\partial_{k_{\alpha_2}}U)\dots (U^\dagger\partial_{k_{\alpha_d}}U)\right] d^{d}k\\ &= i(-2i)^n (2n+1)!\Omega_{2n+1}\text{sign}\left\{ \text{Det}[O] \right\}\\
		& = (-2i\pi)^{n+1} \frac{(2n+1)!}{n!} \text{sign}\left\{ \text{Det}[O] \right\},
		\end{split}
		\end{equation} 
	where $\Omega_n$ is the solid angle of the $n-$sphere. By plugging in the above integral into the winding number definition in Eqs~\eqref{winding_number} and ~\eqref{winding_density}, we prove that
	\begin{equation}
	\nu_{2n+1}[U] = \text{sign}\left\{ \text{Det}[O] \right\} =\pm 1.
	\end{equation}
	We note that this result does not depend explicitly on the form of $O$, i.e. the singularity can be adiabatically deformed without changing the invariant winding number.
	
	\section{Boundary of $\bb{Z}-$indexed AFT}
	In the main text, we derive the boundary signature of the $\bb{Z}_2$-indexed descendant from the the boundary of the $\bb{Z}-$indexed parent for each dimensional hierarchy. In this section, we obtain the boundary behavior of a return map with $\bb{Z}$ classification through direct calculation. This $\bb{Z}$ classification is related to either class A for odd space-time dimensions or class AIII for even space-time dimensions. We first study the former case.
	
	The class A $\bb{Z}$ classification is realized in a return map with even $d=2n+2$ spatial dimensions, its boundary thus has odd $2n+1$ dimensions in space. We assume a generalized ribbon geometry with two surfaces perpendicular to the $d^\text{th}-$dimension. As a result, the return map in this geometry can be approximately partitioned into the boundary and bulk parts as
	\begin{equation}
	\tilde{R}(\kk,t) = \begin{pmatrix}
	\tilde{R}_1(\kk,t) & 0 & 0 \\
	0 & R(\kk,t) & 0\\
	0 & 0 & \tilde{R}_3(\kk,t) 
	\end{pmatrix},
	\end{equation}
	where $\kk=(k_1,k_2,\dots,k_{d-1})^T$. The first and third rows correspond to the upper and lower surface, while the middle row represents the bulk return map that satisfies the periodic condition $R(\kk,t)=R(\kk,t+T)$. We can also define an auxiliary matrix
	\begin{equation}
	Q = \begin{pmatrix}
	0 & 0 & 0\\
	0 & Q_2 & 0\\
	0 & 0 & \mathbbm{1}
	\end{pmatrix}
	\end{equation} 
	to select the anomalous behavior only at the lower boundary. For conciseness, we use the notation $\tilde{M}_\alpha = \tilde{R}^\dagger\partial_{k_{\alpha}}\tilde{R}$ with a derivative identity
	\begin{equation}
	\partial_{k_\beta} \tilde{M}_\alpha =- \tilde{M}_\beta\tilde{M}_\alpha + \tilde{R}^\dagger\partial_{k_\beta k_{\alpha}} \tilde{R},
	\end{equation}
	where the symmetric term can be canceled by adding total derivatives that do not contribute the integral over the BZ, so we only need to keep the anti-symmetry term. By using the same formula for the winding number, we can compute the number of branch-cut-crossing Dirac modes on the boundary,
	\begin{widetext}
		\begin{equation}
		\begin{split}
		\chi_{\text{Dirac}} & = \frac{(-1)^n n!}{(2n+1)!}\left(\frac{i}{2\pi}\right)^{n+1}\int  \epsilon^{\alpha_1\alpha_2\dots\alpha_{2n+1}}\text{Tr}\left[\tilde{M}_{\alpha_1}\tilde{M}_{\alpha_2}\dots \tilde{M}_{\alpha_{2n+1}}.Q\right]d^{2n+1}k\\
		& =  \frac{(-1)^n n!}{(2n+1)!}\left(\frac{i}{2\pi}\right)^{n+1}\int  \epsilon^{\alpha_1\alpha_2\dots\alpha_{2n+1}}\partial_{t}\text{Tr}\left[\tilde{M}_{\alpha_1}\tilde{M}_{\alpha_2}\dots \tilde{M}_{\alpha_{2n+1}}.Q\right]d^{2n+1}kdt\\
		& =  \frac{(-1)^n n!}{2(2n+1)!}\left(\frac{i}{2\pi}\right)^{n+1}\int  \epsilon^{\alpha_0\alpha_1\dots\alpha_{2n+1}}\text{Tr}\left[\tilde{M}_{\alpha_0}\tilde{M}_{\alpha_1}\dots \tilde{M}_{\alpha_{2n+1}}[Q,\tilde{M}_{\alpha_{2n+1}}]\right]d^{2n+2}k
		\end{split}
		\end{equation}
		In the last row, we have identify $t\equiv k_0$ so that $\alpha_{0,\dots,2n+1}\in\{0,\dots,2n+1\}$ and the total derivatives we add to cancel symmetric terms are with respect to the spatial momentum. From the construction of $Q$, the commutation is only non-zero for the bulk (the middle row of $\tilde{R}$) so that we can make the substitution $\tilde{R}\to R$ and accordingly $\tilde{M}\to M$. We also have the following identity from the construction of $Q$ \cite{Rudner2013}
		\begin{equation}
		\text{Tr}\left\{A[Q,B] \right\} = \frac{i}{2\pi}\int dk_{d}A(k_{d})\partial_{k_{d}}B(k_{d}).
		\end{equation}
		The number of boundary Dirac modes is given by
		\begin{equation}
		\begin{split}
		\chi_{Dirac} &= \frac{(-1)^n n!}{2(2n+1)!}\left(\frac{i}{2\pi}\right)^{n+2}\int  \epsilon^{\alpha_0\alpha_1\dots\alpha_{2n+1}}\text{Tr}\left[M_{\alpha_0}M_{\alpha_1}\dots M_{\alpha_{2n}} \left(\partial_{k_{2n+2}}M_{\alpha_{2n+1}}\right)\right]d^{2n+3}k \\
		& =\frac{(-1)^n n!}{2(2n+1)!}\left(\frac{i}{2\pi}\right)^{n+2}\frac{-1}{2n+3}\int \epsilon^{\alpha_0\alpha_1\dots\alpha_{2n+2}}\text{Tr}\left[M_{\alpha_0}M_{\alpha_1}\dots M_{\alpha_{2n+1}} M_{\alpha_{2n+1}}\right]d^{2n+3}k  = \nu_{2n+3}[R]
		\end{split}
		\end{equation}
	\end{widetext}
	We have generalized the proof in Ref.~\cite{Rudner2013} to arbitrary odd space-time dimensions, showing that the $\bb{Z}$ index of class A return maps gives the number of boundary Dirac modes across the branch cut.  On the other hand, as presented in the main text in details, gapless chiral systems can be thought of as an enlarged system with an artificial dimension $\beta$ where the chiral symmetry flips the $\beta$ parameter, thus pinning the nodal point to $\beta=0$. As a result, the $\bb{Z}$ index of class AIII return maps analogously provides the number of even-dimensional chiral Dirac modes on the boundary.
	
\end{document}